\documentclass[fleqn,useAMS,usenatbib]{mn2e}
\usepackage{epsfig}
\usepackage{graphicx}
\usepackage{times}
\usepackage{fixltx2e}
\usepackage{amsmath} 
\usepackage{float}
\usepackage{subfig}
\usepackage[leftcaption]{sidecap}
\usepackage{colordvi}

\newdimen\figurewidth
\def\gtrsim{\mathrel{\hbox{\rlap{\hbox{\lower3pt\hbox{$\sim$}}}\raise2pt\hbox{$>$}}}}
\def\lesssim{\mathrel{\hbox{\rlap{\hbox{\lower3pt\hbox{$\sim$}}}\raise2pt\hbox{$<$}}}}

\RequirePackage[colorlinks=true
,urlcolor=blue
,anchorcolor=blue
,citecolor=blue
,filecolor=blue
,linkcolor=blue
,menucolor=blue
,pagecolor=blue
,linktocpage=true
,pdfproducer=medialab
]{hyperref}

\let\oldhat\hat 
\renewcommand{\vec}[1]{\mathbf{#1}}
\renewcommand{\hat}[1]{\oldhat{\mathbf{#1}}}
\renewcommand{\vec}[1]{\mathbf{#1}}
\renewcommand{\hat}[1]{\oldhat{\mathbf{#1}}}
\newcommand{\mr}[1]{\mathrm{#1}}
\newcommand{\mean}[1]{\langle #1\rangle}
\def\dpa{{\partial}}

\title[Asteroseismic effects in close binary stars]{Asteroseismic effects in close binary stars}
\author[Springer \& Shaviv]{Ofer M. Springer\thanks{E-mail: springer@phys.huji.ac.il} and Nir J. Shaviv\\
\noindent
Racah Institute of Physics, Hebrew University of Jerusalem, Jerusalem 91904, Israel }
\begin{document}


\pagerange{\pageref{firstpage}--\pageref{lastpage}} \pubyear{2013}
\maketitle
\label{firstpage}

\begin{abstract}

Turbulent processes in the convective envelopes of the sun and stars have been shown to be a source of internal acoustic excitations. In single stars, acoustic waves having frequencies below a certain cutoff frequency propagate nearly adiabatically and are effectively trapped below the photosphere where they are internally reflected. This reflection essentially occurs where the local wavelength becomes comparable to the pressure scale height. In close binary stars, the sound speed is a constant on equipotentials, while the pressure scale height, which depends on the local effective gravity, varies on equipotentials and may be much greater near the inner Lagrangian point ($L_1$). As a result, waves reaching the vicinity of $L_1$ may propagate unimpeded into low density regions, where they tend to dissipate quickly due to non-linear and radiative effects. We study the three dimensional propagation and enhanced damping of such waves inside a set of close binary stellar models using a WKB approximation of the acoustic field. We find that these waves can have much higher damping rates in close binaries, compared to their non-binary counterparts. We also find that the relative distribution of acoustic energy density at the visible surface of close binaries develops a ring-like feature at specific acoustic frequencies and binary separations.

\end{abstract}



\section{Introduction}

With the new generation of satellite based photometry, asteroseismology is nearing the state that helioseismology was in several decades ago. The {\sc most} and {\sc wire} missions readily detected the oscillations of giant stars \citep[e.g.,][]{Barban2007,Stello2008}. {\sc corot} and {\sc kepler} could not only detect the oscillations in many more giant stars \citep[e.g.,][]{DeRidder2009}, but also observe the oscillations in solar-like stars as well \citep{Michel2008,Chaplin2010,Chaplin2011}. We can therefore expect asteroseismology to flourish in the coming years, providing constraints on the structure of various stellar objects, as helioseismology did for the sun. For example, asteroseismology of red giants can already be used to distinguish between H-shell burning and He-core burning of red giants \citep{Bedding2011}, or measure the fast rotation of red giant cores \citep{Beck2011}. A wider range of stellar acoustic phenomena is to be expected from the wider range of stellar objects under study in astroseismology, depending on the extent of the stellar convection zone, stellar rotation and tidal distortion, to name a few.

\cite{Saio1981} and later \cite{Mohan1985} developed a perturbation analysis method to determine the small corrections expected in non-radial adiabatic eigenmodes of both rotating and tidally distorted polytropic stars. 

An alternative description to that of eigenmode analysis, discussed by \cite{Gough1993,Swisdak1999,Gough2007}, is that of ray dynamics, where the so called WKB approximation is applied to acoustic waves in the ``optical limit".  \cite{Lignieres2006} and \cite{Reese2006} used this ray dynamic description to study rotationally distorted stars. They have later shown that there are two different families of rays. One family is of ``normal" eigenmodes which are similar to the non-rotating case, albeit with a reduced degeneracy. The second family is chaotic and therefore describes waves which roam at least part of the star  without forming a standing wave \citep{Lignieres2008}. 

Here we shall consider a related problem, which is that of asteroseismology of tidally distorted stars in binary systems, the most distorted of which fill their Roche lobe completely. Because of the complex geometry, we will follow the ray dynamic description, which is necessary for two primary reasons. First, as was shown by \cite{Lignieres2008} for rotationally distorted stars, we can expect chaotic behaviour which cannot be described by the standard eigenmode analysis. Secondly, we will see in the present analysis that the weaker gravity present in the vicinity of the $L_1$ point gives rise to strong wave dissipation. This phenomenon, where the waves have a significantly reduced lifetime due to localized dissipation, can be more easily described using ray dynamics. 

We begin in \S\ref{sec:stellarModels} by describing the stellar models we used as the background medium. These are elaborated on in Appendix \ref{sec:structure}. In \S\ref{sec:propagation} we describe the propagation of acoustic wave packets and in \S\ref{sec:excitationAndDamping} we calculate their excitation and damping. The results are then summarized in \S\ref{sec:results} and discussed in \S\ref{sec:discussion}.


\section{Binary stellar models}
\label{sec:stellarModels}

Although we are primarily interested in the propagation of acoustic waves, we require stellar models for their background. For this purpose we developed models describing tidally and rotationally distorted binary systems.

These models utilize the Kippenhahn-Thomas (KT) method which modifies the standard spherical solutions using correction factors describing the geometry of the Roche equipotentials. The full description of this modelling is found in Appendix \ref{sec:structure}.

In the following work, we considered binary systems composed of $1M_\odot$ ZAMS stars having solar composition. We limit ourselves to such stars to facilitate their comparison with helioseismological results. 
The resulting models are denoted as $S1$-$S13$. $S1$ is a spherically symmetric model, while $S2$-$S13$ are progressively closer systems, with $S13$ filling its Roche lobe completely. See the table in \S\ref{sec:resultingStellarModels}  for their specific parameters. 

\section{Propagation and trapping of acoustic waves}
\label{sec:propagation}

We now proceed to study the way in which acoustic waves propagate in binaries, postponing a discussion of their excitation and damping mechanisms to the following section. The background stellar structure affects the propagation of acoustic waves in essentially two ways. Near the surface, outgoing waves are reflected back due to a steep stratification (e.g., where stellar scale heights are shorter than the wavelength). Towards the stellar core, the rise in temperature leads to a rise in the adiabatic sound speed $a$, which in turn, causes ingoing non-radial waves to bend outwards. This results in an inner turning point of the waves. In the spherical case, this trapping acts as a resonant cavity, selecting discrete modes of oscillation, known as p-modes in helioseismology. Lacking spherical symmetry in the binary case, we choose to describe the acoustic field, not as a set of global eigenmodes, but instead as a collection of localized wave packets in the WKB (or ray) approximation. In \S\ref{sec:excitationAndDamping}, this also allows us to describe wave excitation and damping as an emission and absorption process of wave packets.

\subsection{Wave equation and local dispersion relation}
\label{sec:waveEquation}

To find the dynamics of a small perturbation of the steady state structure, we express each local quantity as $f(\vec{r},t) = f_0(\vec{r}) + f_1(\vec{r},t)$, where $f_0$ is the static solution (or background), and $f_1$ is a perturbation. We assume that the structural perturbations are small compared with their background values, so that e.g., $\vert \rho_1/\rho_0\vert\ll 1$, $\vert P_1/P_0\vert\ll 1$, and that the resulting velocity field $\vec{v}=\vec{v_1}$, obeys $\Vert \vec{v_1}\Vert\ll a$. We base the following derivation on the conservation equations for mass, momentum, and energy,
\begin{subequations}\begin{align}
\frac{\dpa\rho}{\dpa t} &= -\nabla\cdot(\rho\vec{v}),\label{eq:propContinuity}\\
\rho\frac{d\vec{v}}{dt} &= \rho\vec{g} - \nabla P,\label{eq:propEuler}\\
\rho\frac{dq}{dt} &= \frac{1}{\Gamma_3-1}\left(\frac{dP}{dt} - a^2\frac{d\rho}{dt} \right),\label{eq:propEnergyConservation}
\end{align}\end{subequations}
where $a \equiv \sqrt{\Gamma_1 P / \rho}$, and the adiabatic exponents $\Gamma_1$ and $\Gamma_3$ are defined for an adiabatic process by the relations $P \rho^{-\Gamma_1} = \mr{const}$ and $T \rho^{1-\Gamma_3} = \mr{const}$ respectively. Retaining in (\ref{eq:propContinuity}) and (\ref{eq:propEuler}) terms which are first order in the perturbation, we have that
\begin{subequations}\begin{align}
\frac{\dpa\rho_1}{\dpa t} &= -\nabla\cdot(\rho_0\vec{v}_1),\label{eq:propContinuityLin}\\
\rho_0\frac{\dpa\vec{v_1}}{\dpa t} &= \rho_1\vec{g} - \nabla P_1,\label{eq:propEulerLin}
\end{align}\end{subequations}
where in (\ref{eq:propEulerLin}) we used the hydrostatic equation for the background, $\nabla P_0 = \rho_0 \vec{g}$, and the relation $d\vec{v}_1/dt = \dpa\vec{v}_1/\dpa t$, which holds to first order in $\vec{v}_1$. We also neglect any dependence $\vec{g}$ may have on the perturbation \citep[this is the Cowling approximation which is justified in the high frequency limit we will ultimately focus on; see][]{Cowling1941}. Using the definition of the convective derivative, and restricting ourselves to adiabatic waves, for which we assume $dq/dt=0$, we obtain the linearized version of (\ref{eq:propEnergyConservation}),
\begin{equation}
\frac{\dpa P_1}{\dpa t} = \label{eq:propEnergyConservationLin}
-\vec{v}_1\cdot\nabla P_0 - \Gamma_{1,0} P_0 \nabla\cdot\vec{v}_1.
\end{equation}
Treating the medium as locally plane-parallel, \citet{Gough1993} obtains from the linearized conservation equations (\ref{eq:propContinuityLin}), (\ref{eq:propEulerLin}), and (\ref{eq:propEnergyConservationLin}), the following wave equation 
\begin{equation}\label{eq:propGough1}
\left(\frac{\dpa^2}{\dpa t^2} + \omega_\mr{c}^2\right)
\frac{\dpa^2\Psi}{\dpa t^2} -
a^2\frac{\dpa^2}{\dpa t^2}\nabla^2\Psi - 
a^2N^2\nabla_\mr{h}^2\Psi = 0
\end{equation}
for the scalar quantity $\Psi=\delta P/\rho_0^{1/2}$. Here $\delta P$ is the Lagrangian pressure perturbation, $N$ is the Brunt-V\"ais\"al\"a buoyancy frequency, $\nabla_\mr{h}^2$ is the horizontal Laplacian operator, $\omega_\mr{c}$ is defined by
\begin{equation}\label{eq:propOmegaC}
\omega_\mr{c}^2 \equiv \frac{a^2}{4H_\rho^2}\left(1-2\hat{n}\cdot\nabla H_\rho\right),
\end{equation}
and $H_\rho \equiv -\rho_0 (d\rho_0/dn)^{-1}$ is the density scale height. We use $n$ here to denote the vertical coordinate. In the high frequency limit $\omega^2\gg N^2$, which is applicable in the convection zone where stratification is nearly adiabatic. Gough sets $N=0$ in (\ref{eq:propGough1}) and obtains 
\begin{equation}\label{eq:propGough2}
\left(\frac{\dpa^2}{\dpa t^2} + \omega_\mr{c}^2\right)\Psi -
a^2\nabla^2\Psi = 0.
\end{equation}
Note that $N={\cal O}(\omega)$ at the outer turning point, however the horizontal wavenumber of low $\ell$-degree waves is much smaller than the vertical wavenumber there, so that $\nabla_\mr{h}^2\Psi \ll \nabla^2\Psi$, and the third term in Eq.\,(\ref{eq:propGough1}) may still be neglected. 

\begin{figure}
\includegraphics[width=0.5\textwidth]{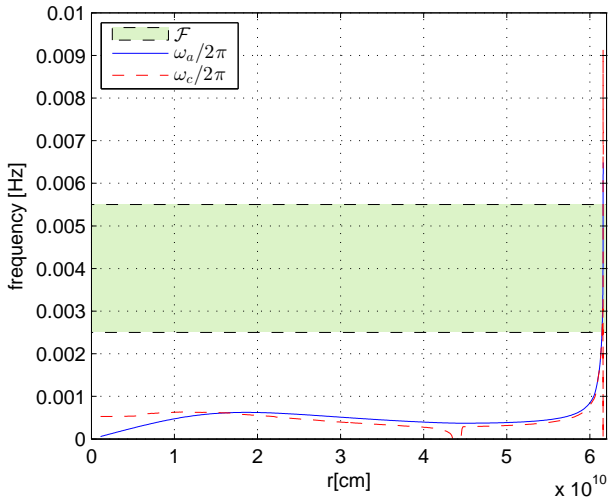}
\caption{Radial profile of frequencies $\omega_\mr{a}$, $\omega_\mr{c}$, and frequency range ${\cal F}$ throughout model $S1$.}
\label{fig:omegaAomegaCfull}
\end{figure}

\begin{figure}
\includegraphics[width=0.5\textwidth]{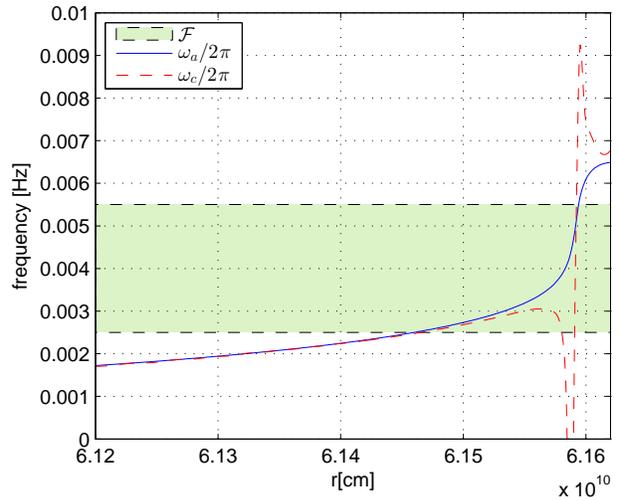}
\caption{Radial profile of frequencies $\omega_\mr{a}$, $\omega_\mr{c}$, and frequency range ${\cal F}$ near the surface of model $S1$.}
\label{fig:omegaAomegaCouter}
\end{figure}

To derive a local dispersion relation and as a preparation for the wave packet description of the following section, we look for solutions to (\ref{eq:propGough2}) which can locally be considered planar. First, we note that in a homogeneous medium we would have that $a=\mr{const}$, $\omega_\mr{c}=0$, and a solution to (\ref{eq:propGough2}) would be given by a superposition of plane waves of the form
\begin{equation}\label{eq:propGoughPlane}
\Psi(\vec{r},t) = \Psi_0e^{i(\omega t - \vec{k}\cdot\vec{r})}.
\end{equation}
In a stratified medium, we write the general solution in the form
\begin{equation}\label{eq:propGoughAnsatz}
\Psi(\vec{r},t) = \Psi_0(\vec{r},t)e^{i\Lambda\Phi(\vec{r},t)},
\end{equation}
where we assume that the background quantities $\omega_\mr{c}$ and $a$, as well as the wave amplitude and phase functions $\Psi_0$ and $\Phi$, all vary on a length scale much larger than the local wavelength of the solution. Using $\Lambda$ to denote this length scale ratio, and treating $\Lambda^{-1}$ as a small parameter, we substitute the ansatz (\ref{eq:propGoughAnsatz}) into the wave equation (\ref{eq:propGough2}). Keeping leading order terms in $\Lambda^{-1}$, we find
\begin{equation}\label{eq:propGoughLeadingOrder}
\left(\frac{\dpa \Phi}{\dpa t}\right)^2 - 
\left(\frac{\omega_\mr{c}}{\Lambda}\right)^2 - 
a^2\left(\nabla\Phi\right)^2 = 0,
\end{equation}
where we have taken into account the possibility that $\omega_\mr{c}={\cal O}(\Lambda)$ near the surface. From an analogy between the plane wave solution (\ref{eq:propGoughPlane}) and the general solution (\ref{eq:propGoughAnsatz}), we define the local frequency and wave vector of a wave packet located at position $\vec{r}$ at time $t$ as
\begin{align}\label{eq:propPacket}
\omega \equiv \Lambda \frac{\dpa \Phi}{\dpa t}, ~~~\vec{k} \equiv -\Lambda \nabla\Phi.
\end{align}
Using these quantities, we can now express (\ref{eq:propGoughLeadingOrder}) in the form of a local dispersion relation,
\begin{equation}
\omega^2 = \omega_\mr{c}^2 + a^2\vec{k}^2.\label{eq:propAcousticDispersionNonisothermal}
\end{equation}
In an isothermal medium, the pressure and density scale heights are constant and equal to one another \citep[see][Ch.\,5]{Mihalas1984}, so that in this limit $\omega_\mr{c}$ reduces to $\omega_\mr{a} \equiv a/2H_P = \gamma g/2a$, with $\gamma \equiv \Gamma_{1,0}$. Computing $\omega_\mr{c}$ and $\omega_\mr{a}$ in $S1$ (see Figs.\,\ref{fig:omegaAomegaCfull} and \ref{fig:omegaAomegaCouter}), we find that they are comparable in the interior of the envelope, and reach a similar value at the upper atmosphere. There, these frequencies essentially differ in the superadiabatic layer, where $\Gamma_1$ changes rapidly \citep[cf.,][Fig.\,7.5]{JCD2003}.
In \S\ref{sec:propRays}, using the WKB approximation, we will see that such acoustic waves reflect inwards once they reach a surface layer where $\omega\approx\omega_\mr{c}$. We see then that by using the isothermal dispersion relation 
\begin{equation}
\omega^2 = \omega_\mr{a}^2 + a^2\vec{k}^2 \label{eq:propAcousticDispersion}
\end{equation}
instead of (\ref{eq:propAcousticDispersionNonisothermal}), we may be underestimating the maximal trapped frequency by approximately $30\%$. In the following analysis, relation (\ref{eq:propAcousticDispersion}) was preferred due to numerical stability issues that arise when evaluating (\ref{eq:propAcousticDispersionNonisothermal}). Notice that the term $\omega_\mr{c}^2$ in (\ref{eq:propAcousticDispersionNonisothermal}) contains a second derivative of $\rho_0$ with respect to $n$.

We now focus our attention on waves in the frequency range ${\cal F}=\left[2.5,5.5\right]~\mr{mHz}$, which roughly overlaps the solar acoustic excitation spectrum \citep[see][and Ch.\,\ref{sec:excitationAndDamping}]{LibbrechtWoodard1991}. These frequencies lie above $\nu_\mr{a}\equiv\omega_\mr{a}/2\pi$ throughout most of $S1$'s profile (see Fig.\,\ref{fig:omegaAomegaCfull}). Near the surface, $\nu_\mr{a}$ rises steeply, reaching a maximum value of $\nu_\mr{a}^\mr{ph} \approx 6.5~\mr{mHz}$ at the photosphere. 

The photospheric acoustic cutoff frequency $\omega_\mr{a}^\mr{ph} \equiv 2\pi\nu_\mr{a}^\mr{ph}$ sets the maximum frequency of trapped acoustic waves. Waves having frequencies above $\omega_\mr{a}^\mr{ph}$ are not reflected at the surface, but instead propagate into the upper atmosphere, where their damping is expected to be much greater. We assume that such waves are absorbed there (see \S\ref{sec:excitationAndDamping}). In binaries, the fact that  $g$ varies on equipotentials causes $\omega_\mr{a} = \gamma g/2a$ to vary in the photosphere as well, affecting the trapping mechanism just described (this is elaborated on in \S\ref{sec:propTrapping}).

In the spherical case, these acoustic waves are the constituents of the p-modes defined in the eigenmode description of stellar oscillations. In this description, a Lagrangian perturbation $\delta f(\vec{r},t)$, of a steady-state quantity $f(\vec{r})$, is expressed as a combination of modes, each having an angular dependence proportional to a spherical harmonic $Y_\ell^m(\theta,\phi)$. The inner turning point $r_t$, and the $\ell$-degree of a p-mode are related by \citep[see][]{JCD2003}
\begin{equation}
\frac{a^2(r_\mr{t})}{r_\mr{t}^2} = \frac{\omega^2}{\ell(\ell+1)}.\label{eq:propModelL}
\end{equation}
In the sun, the inner turning point of acoustic waves having frequencies in ${\cal F}$, and $\ell \approx 50$, confines them to the convective envelope \citep[cf.,][]{LibbrechtWoodard1991}. In the following WKB approximation we find this is also the case in $S1$. Throughout the rest of this section, we focus on these waves for the following three reasons. First, in the envelope, the previously employed plane-parallel assumption is a reasonable one. Secondly, in this low $\ell$-degree regime, dropping the the third term in Eq.\,(\ref{eq:propGough1}) is justified. Finally, this is also a practical choice which allows us to reduce the parameter space that should subsequently be studied.
\subsection{Propagation of localized wave packets}\label{sec:propRays}
Using expressions (\ref{eq:propPacket}) for the frequency and wave vector of a wave packet, and expressing its group velocity, we find the following equations of motion for the components of the packet's wave vector and position 
\begin{align}\label{eq:propPacketEOM}
\frac{dk_i}{dt} = -\frac{\dpa\omega}{\dpa r_i}, ~~~
\frac{dr_i}{dt} = \frac{\dpa\omega}{\dpa k_i},
\end{align}
where $\omega = \omega(\vec{r},\vec{k})$ depends on the position and wave vector through (\ref{eq:propAcousticDispersion}). Treating $r_i$ and $k_i$ as canonical coordinates and conjugate momenta respectively, then Eqs.\,(\ref{eq:propPacketEOM}) are Hamilton's equations for the Hamiltonian $\omega(\vec{r},\vec{k})$. Expressing the differential of $\omega$ with respect to time, following the packet, we have that
\begin{equation}\label{eq:propPacketDOmega}
\frac{d\omega}{dt} = \frac{\dpa\omega}{\dpa t} + \frac{\dpa\omega}{\dpa r_i}\frac{dr_i}{dt} + \frac{\dpa\omega}{\dpa k_i}\frac{dk_i}{dt},
\end{equation}
where repeated indices are summed over. For a given $\vec{r}$ and $\vec{k}$, $\omega$ depends on the static background structure (to first order in the perturbation), so that $\dpa\omega/\dpa t = 0$. Using also the equations of motion (\ref{eq:propPacketEOM}) in (\ref{eq:propPacketDOmega}) we find that $d\omega/dt=0$, so that the packet frequency $\omega$ remains constant during its propagation (this is equivalent to the energy being a constant of motion in systems having a time-independent Hamiltonian).

Working in the previously defined plane parallel coordinate system, we can now find the position of the outer turning point of the wave packets. Setting $dr_n/dt = 0$ in (\ref{eq:propPacketEOM}) and using (\ref{eq:propAcousticDispersion}) we find
\begin{equation}
0 = \frac{\dpa\omega}{\dpa k_n} = \frac{a^2 k_n}{\omega},
\end{equation}
so that $k_n$ vanishes at the outer turning point. Setting $k_n=0$ in the dispersion relation, we find that this occurs at a layer in which
\begin{equation}
\omega_\mr{a}^2(n) = \omega^2 - a^2k_h^2,
\end{equation}
where $k_h^2$ is the horizontal component of the wave vector. Assuming that near the surface $(ak_h/\omega)^2\ll 1$ (cf., \S\ref{sec:waveEquation}), we see that the reflection occurs where the acoustic cutoff frequency is approximately equal to the packet's frequency $\omega_\mr{a}(n)\approx\omega$, with this condition being exact for vertically propagating packets. In general, the path of a wave packet may be determined by integrating the ODEs (\ref{eq:propPacketEOM}) starting with some initial values for $\vec{r}_0$ and $\vec{k}_0$. The packet's dispersion relation (\ref{eq:propAcousticDispersion}) depends in a non-trivial way on the background stellar structure, through $a(\vec{r})$ and $\omega_\mr{a}(\vec{r})$. We thus integrate (\ref{eq:propPacketEOM}) numerically. This is performed using the standard fourth order Runge-Kutta method \citep[see e.g.,][Ch.\,16]{PressTeukolsky1992}, with the accumulated error in the packet frequency $\omega$ being less than $0.5\%$ (recall that $\omega$ should ideally be conserved along the path).

\begin{figure}
\includegraphics[width=0.5\textwidth]{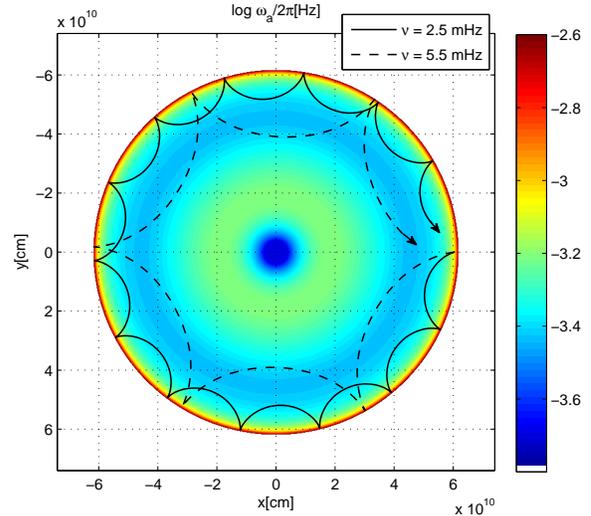}
\caption{Paths of two trapped wave packets propagating in a plane through model $S1$'s centre. Both packets have inner turning points corresponding to $\ell=50$.}
\label{fig:rayPathsS1}
\end{figure}

In single non-rotating stars, assuming spherical symmetry, the path of a given packet is confined to a plane defined by its initial position $\vec{r}_0$, the direction of its initial wave vector $\hat{\vec{k}}_0$, and the position of the stellar centre (in binaries this is generally not the case). In Fig.\,\ref{fig:rayPathsS1}, we show the path of two wave packets, having different frequencies in ${\cal F}$, propagating between their inner and outer turning points in model $S1$.
The initial conditions of the two packets have been chosen so that their inner turning points correspond to $\ell = 50$, with the $\ell$-degree defined by (\ref{eq:propModelL}). To aid in setting up these initial conditions, $\theta(\omega)$, the dependence of the angle $\cos(\theta)=\hat{\vec{k}}_0\cdot(-\hat{\vec{r}})$ on the packet frequency $\omega$ for $\ell = 50$, was computed at the arbitrary stellar layer $T=10^5 ~\mr{K}$.

The wave packets of Fig.\,\ref{fig:rayPathsS1} were then initialised in the following way. Their initial position $\vec{r}_0$ was chosen so that in $S1$, $T(\vec{r}_0) = 10^5 ~\mr{K}$, and the angle $\theta$ between the direction of their initial wave vector $\hat{\vec{k}}_0$ and $-\hat{\vec{r}}$ was chosen according to the aforementioned $\theta(\omega)$ relation. Finally, using dispersion relation (\ref{eq:propAcousticDispersion}), their wavenumber was set according to $\vec{k}_0^2=[\omega^2-\omega_\mr{a}^2(\vec{r}_0)]/a^2(\vec{r}_0)$.

\subsection{Effects of binary structure on trapping}\label{sec:propTrapping}

In binaries, due to the variation of $g$ on equipotentials and the fact that $\omega_\mr{a} \propto g$, we expect the aforementioned wave trapping mechanism to be significantly modified. 

\begin{figure*}
\centering
\subfloat[In the $x$-$y$ plane]{\label{fig:omegaAofThetaXY}\includegraphics[width=0.4\textwidth]{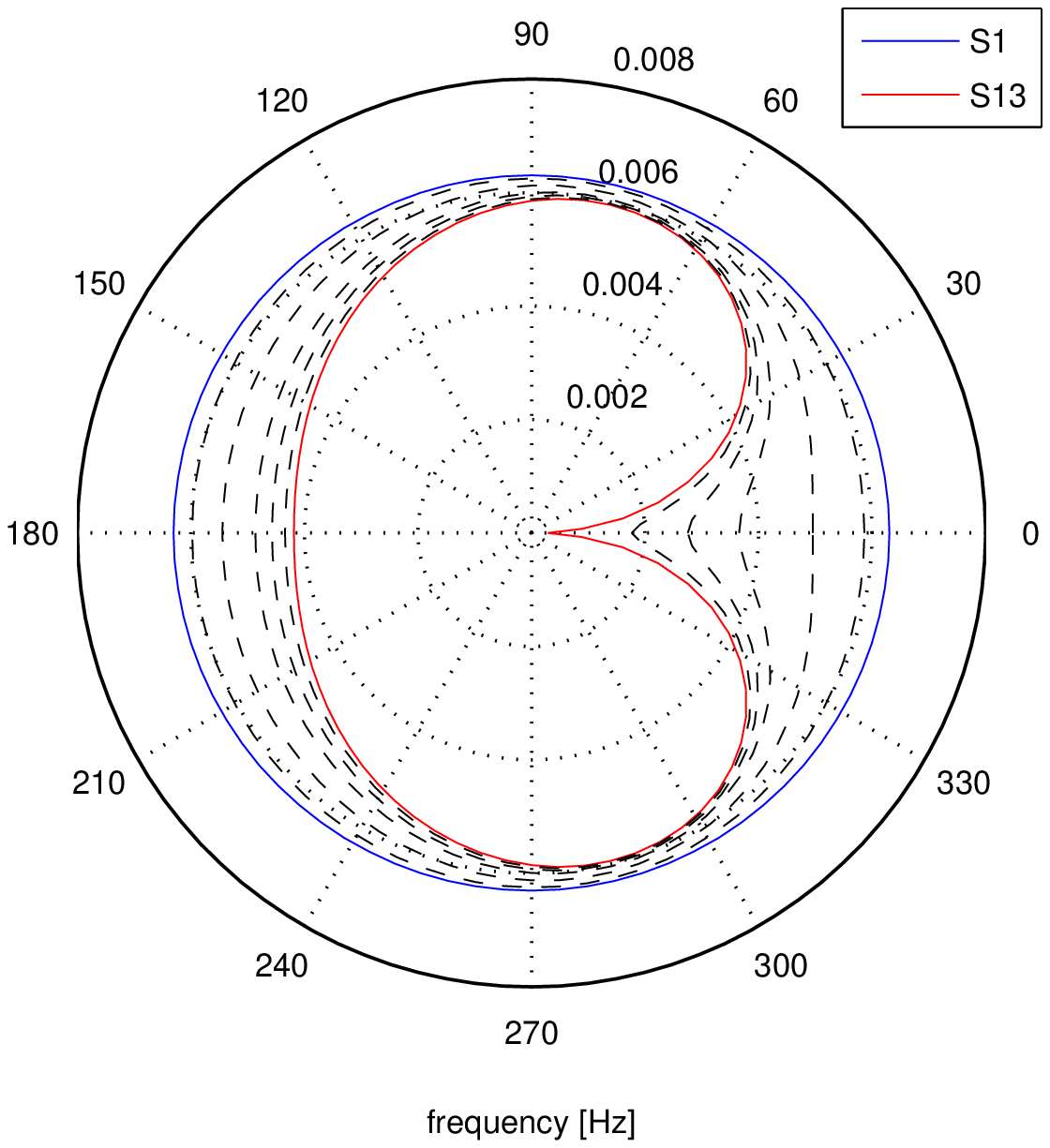}}
\subfloat[In the $x$-$z$ plane]{\label{fig:omegaAofThetaXZ}\includegraphics[width=0.4\textwidth]{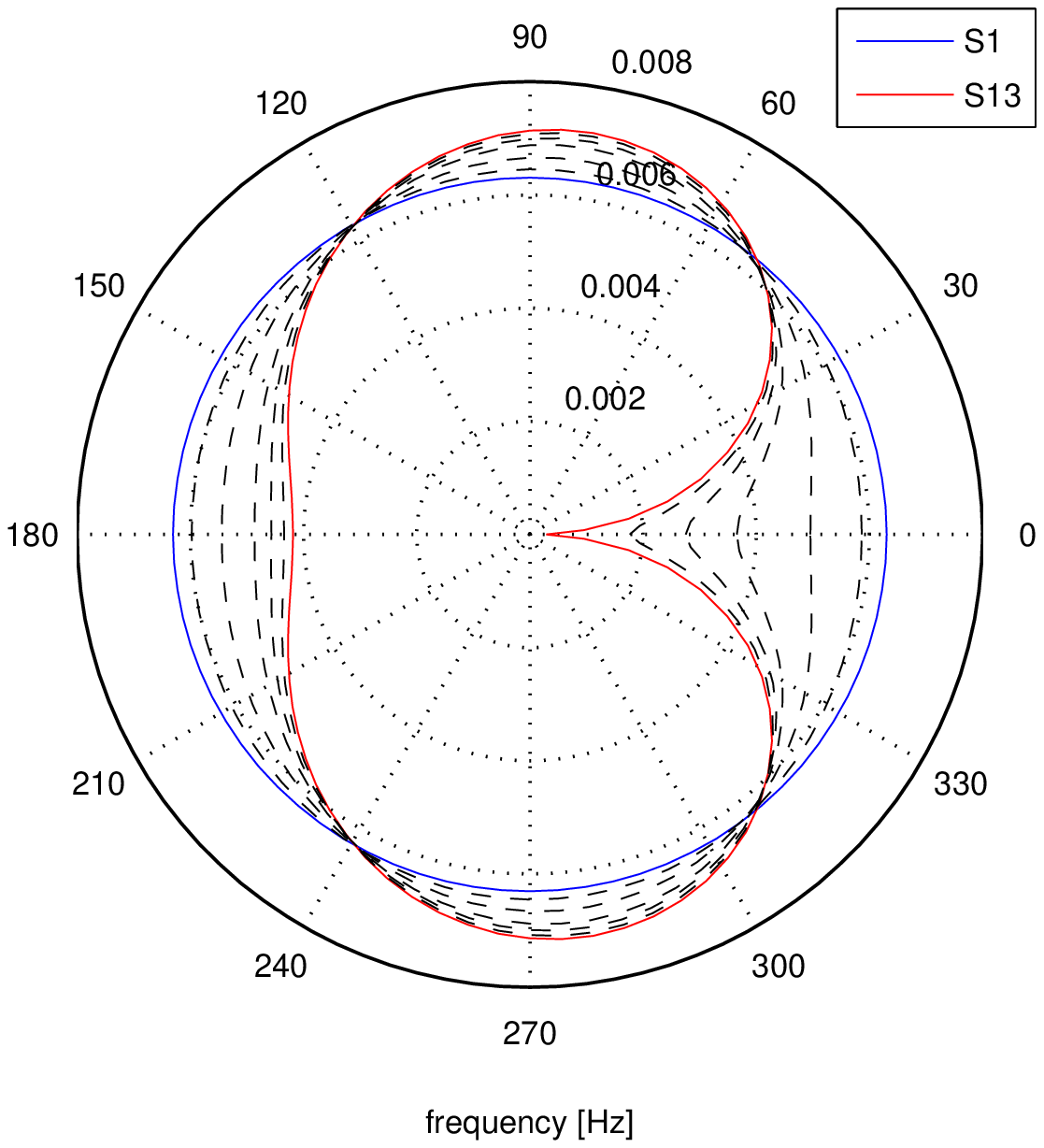}}
\caption{Polar plots of $\omega_\mr{a}^\mr{ph}/2\pi$ in models $S1$, $S3$, $S5$, $S7$, $S9$, $S11$, and $S13$. The polar coordinate is the angle $\theta$ between the positive $x$ axis and a ray originating at the stellar centre, in either the $x$-$y$ or the $x$-$z$ planes, with $\theta=0$ thus pointing towards $L_1$. See Fig.\,\ref{fig:titleWithAxes} for a full definition of the coordinates used here and throughout.}
\label{fig:omegaAofTheta}
\end{figure*}

\begin{figure}
\includegraphics[width=0.5\textwidth]{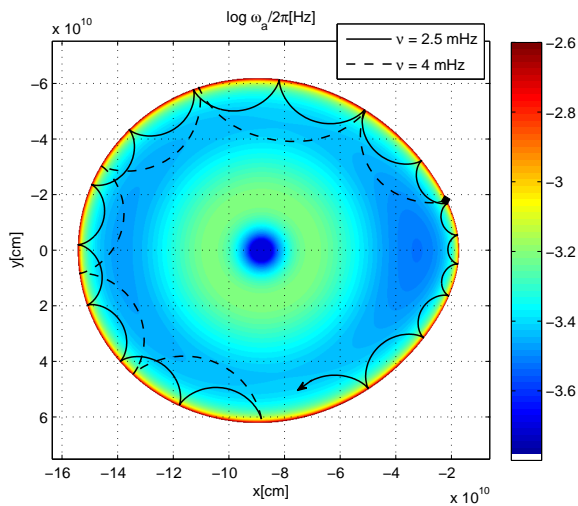}
\caption{Path of two packets in the $x$-$y$ plane of $S9$. The higher frequency packet is absorbed near $L_1$ (indicated with a diamond), while the lower frequency packet propagates through this region.}
\label{fig:rayPathsS9}
\end{figure}

\begin{figure}
\includegraphics[width=0.5\textwidth]{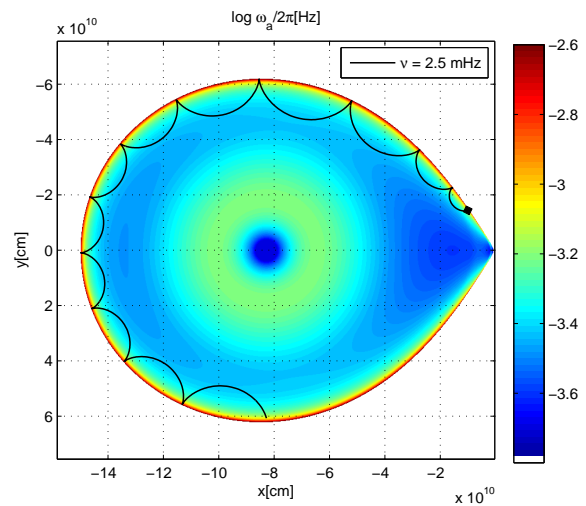}
\caption{Path of a wave packet in the $x$-$y$ plane of $S13$. The packet is eventually absorbed near $L_1$ (indicated with a diamond).}
\label{fig:rayPathsS13}
\end{figure}

In Fig.\,\ref{fig:omegaAofTheta} we show, $\omega_\mr{a}^\mr{ph}$, the value of $\omega_\mr{a}$ at the photospheric layer $T=5000~\mr{K}$, in some of the stellar models of \S\ref{sec:structure}. The coordinate system used in the following is defined in Fig.\,\ref{fig:titleWithAxes}. We see that as the binary separation parameter $d$ decreases (with increasing model index; see Table \ref{tbl:modelResultingParameters}), the values of $\omega_\mr{a}^\mr{ph}$ depart from the corresponding spherical model ($S1$) values, with $\omega_\mr{a}^\mr{ph}$ almost vanishing near $L_1$ in $S13$. This is simply due to the reduced effective gravity there.
The internal $x$-$y$ plane values of $\omega_\mr{a}$ in models $S1$, $S9$, and $S13$ are also shown in Figs.\,\ref{fig:rayPathsS1}, \ref{fig:rayPathsS9}, and \ref{fig:rayPathsS13}. We therefore see that in the binary models, the surface reflection of wave packets depends on the region the packet impinges on. In models $S1$-$S12$ there exists a non-zero frequency $\omega_\mr{a}^\mr{ph,min}$ below which packets of all frequencies are trapped. In $S13$ on the other hand, practically all frequencies are greater than $\omega_\mr{a}^\mr{ph,min}$, hence all packets reaching the vicinity of $\phi=0$ (see Fig.\,\ref{fig:omegaAofTheta}) are absorbed there. In Figs.\,\ref{fig:rayPathsS9} and \ref{fig:rayPathsS13} we show some wave packets propagating in models $S9$ and $S13$, with some of them being absorbed near $L_1$.

Although an $\ell$-degree isn't strictly defined in binaries, we can still confine packets to the binary envelope by initialising the packet positions to the layer $T=10^5 ~\mr{K}$ and their wave vectors according to the $\theta(\omega)$ relation suitable for $\ell=50$ in $S1$ (see \S\ref{sec:propRays}). This is the way in which packets were initialised in Figs.\,\ref{fig:rayPathsS9} and \ref{fig:rayPathsS13}. 

We will see in \S\ref{sec:results} that although most packets having frequencies $\omega>\omega_\mr{a}^\mr{ph,min}$ are quickly absorbed (within a few stellar sound crossing times), a small fraction of packets may continue to propagate for much longer durations on a path for which $\omega<\omega_\mr{a}^\mr{ph}$ at the surface. Such long-lived paths can exist within a band around the $x$ axis of the binary models (see Fig.\,\ref{fig:titleWithAxes} showing such a packet propagating in $S13$). This region corresponds in Fig.\,\ref{fig:omegaAofTheta} to the two $\omega_\mr{a}^\mr{ph}$ bulges visible at $\phi=90$ and $\phi=270$. Although they may be few, we will show that these packets significantly contribute to the acoustic energy density in binaries due to their relatively long lifetimes. 


\section{Excitation and damping of acoustic waves}\label{sec:excitationAndDamping}

We wish to evaluate how binary acoustic energetics is affected by the reduced wave trapping presented in the previous section. In this work, our focus is on solar-like oscillations. These oscillations are presently thought to be excited by near surface turbulent convection, although the exact mechanisms of both their excitation and damping are not yet fully understood \citep[see][for a review of the present state of this field]{Samadi2009}. While large uncertainties exist in currently available theoretical estimates of asteroseismic damping rates, we have reasons to believe that the reduced trapping in binaries has a significant energetic effect. In the sun, globally coherent oscillations have been observed to have lifetimes between days to months \citep[cf.,][]{LibbrechtWoodard1991}. Modes having comparable lifetimes, inferred from the widths of the resonant peaks in the doppler power spectrum, have also been detected in red giants \citep{DeRidder2009}. We will see in the following section that the reduced trapping in binaries may shorten wave packet lifetimes to the sound crossing time-scale, which is of the order of minutes to hours, and it is therefore expected in certain circumstances to be the dominant damping process.

\subsection{Wave packet emission}\label{sec:emission}

We do not attempt to model the frequency and angular dependence of the acoustic emission rate, $f(\vec{r},\hat{\Omega},\nu)$, having units of power per unit volume, per unit solid angle, per unit frequency. This function is expected to depend on the details of the turbulent convection process, e.g., through its spatial and temporal spectra (cf., \citealt{Samadi2003}, \citealt{Chaplin2005}). We only assume that emission occurs primarily in the near surface superadiabatic layer \citep[cf.,][]{Osaki1990}, that it occurs at a constant rate per unit surface area (in accordance with the way we model the superadiabatic layer in \S\ref{sec:ktCorrections}), and that this layer can be considered locally plane parallel, with the emission rate possibly depending on the angle $\cos(\theta) \equiv \hat{\Omega}\cdot\hat{g}$.

\subsection{Asteroseismic damping rates}

\begin{figure}
\includegraphics[width=0.5\textwidth]{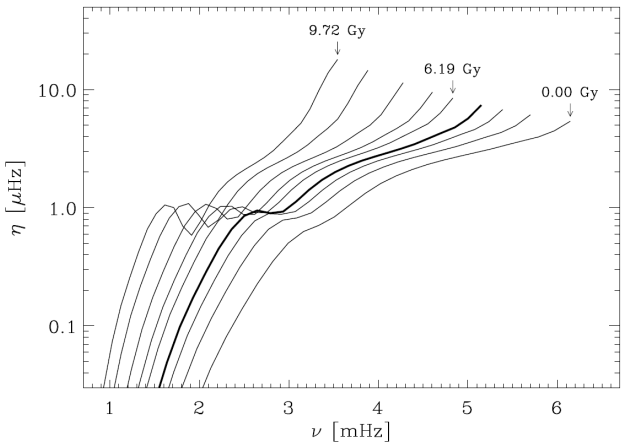}
\caption{Damping rates as a function of mode frequency in an evolving $1\mr{M}_\odot$ stellar model (see text). The thick curve is the result for the age of the present day sun. Taken from \protect\cite{Houdek1999}. }
\label{fig:houdek}
\end{figure}

A large number of processes are currently suspected to contribute to the damping of solar-like oscillations. Of these processes, those presently considered most significant are the radiative damping of the waves in the atmospheric layers, whereby energy in the wave is transferred from compressed to rarefied regions by radiation (cf., \citealt{Mihalas1984}, Ch.\,9, and \citealt{JCD1983}), and the interaction of the turbulence with the oscillation. This interaction is usually modelled by first separately solving the linear oscillation and the convection, using e.g. mixing-length theory (MLT), and then calculating their coupling. In turn, this coupling is generally separated into a direct mechanical one (through the momentum conservation equation), and a modulation of the convective heat flux by the oscillation (through the energy conservation equation) \citep[cf.,][]{Gough1977}. Using stellar models constructed with a modified MLT description of convection, and accounting for the above damping processes in the non-adiabatic oscillation problem, \cite{Houdek1999} estimated the damping rates of radial modes ($\ell=0$) for an evolving $1\mr{M}_\odot$ model (see Fig.\,\ref{fig:houdek}).
The damping rates $\eta(\nu)$ in fig.\,\ref{fig:houdek} are the imaginary parts of the computed complex eigenfrequencies $\omega=\omega_r+i\eta$ of modal perturbation quantities (such as $\vec{v}_1$). The total mode energy $E=\int\rho_0 \vec{v}_1^2 d^3r$ (in an undriven mode) thus evolves as $dE/dt = -2\eta E$. According to \cite{LibbrechtWoodard1991}, the observed solar damping rates of low $\ell$-degree modes depend little on the value of $\ell$ (with $\eta_{\ell=50}$ being larger than $\eta_{\ell=0}$ by approximately $30\%$), while they vary by about two orders of magnitude in the frequency range ${\cal F}$. We will therefore use in the following section the zero-age damping rates of Fig.\,\ref{fig:houdek} to estimate the lifetimes of non-radially propagating packets in our models.

\subsection{Wave packet lifetimes}\label{sec:lifetimes}

Lacking a reflection mechanism for packets having frequencies $\omega > \omega_\mr{a}^\mr{ph}$ in our model, we expect such packets to propagate upwards through the steadily decreasing density profile of the stellar atmosphere. In our model photosphere (described in \S\ref{sec:photosphere}), the density drops by more than two orders of magnitude, while the temperature converges to a finite value of approximately $5000~\mr{K}$. The adiabatic sound speed also keeps an approximately constant value throughout the photosphere. Consequently, a wave packet entering the base of the photosphere is expected to maintain its spatial dimensions $\lambda$. As $\rho_0$ decreases, the packet velocity amplitude $v_{1,\mr{p}}$ must increase to conserve its total (kinetic plus internal) energy $E_\mr{p} \approx \lambda^3 \rho_0 v_{1,\mr{p}}^2$. Assuming that $\rho_0\rightarrow 0$ in the upper atmosphere, then the increased packet velocity amplitude may become non-negligible compared to the local sound speed. This results in a non-linear evolution of the wave (e.g., the adiabatic sound speed will considerably vary within the wave), to a steepening of the wave profile, and to the eventual formation of quickly dissipating shock fronts \citep[see e.g.][in which the effects of radiative damping are also considered]{Ulmschneider1971}. This atmospheric heating mechanism, suggested by \cite{Schwarzschild1948}, is presently considered to account for most of the non-radiative heating of the lower solar chromosphere \citep[cf.,][]{Ruderman2006}. The possibility that other reflection mechanisms, not modelled in our work, may be present in a more realistic stellar atmosphere \citep[e.g.,][]{BalmforthGough1990} is addressed in \S\ref{sec:discussion}.

We therefore treat wave packet damping in the following simplified way:
\begin{itemize}
\item Assuming that throughout the surface regions from which a packet is reflected, wave damping conditions in our models may be approximated using the rates of \cite{Houdek1999}, then the maximal effective lifetime of packets, having frequency $\nu$, is defined to be $t_\mr{max} \equiv 1/2\eta(\nu)$.
\item We assume that packets reaching the upper photosphere are immediately absorbed. We therefore define a packet's effective lifetime to be $t \equiv \min(t_\mr{max}, t_\mr{abs})$, where $t_\mr{abs}$ is the time elapsed between packet emission and absorption.
\end{itemize}


\section{Results}\label{sec:results}

Using the wave packet equations of motion derived in \S\ref{sec:propRays}, and based on the assumed acoustic emission rates and packet lifetimes described in \S\ref{sec:emission} and \S\ref{sec:lifetimes}, we studied the acoustic energetics of our binary models, specifically their mean packet lifetimes and their relative surface acoustic energy densities. This was computed in each of the binary models ($S2$-$S13$), for $13$ equi-spaced values of the wave frequency in the range ${\cal F}=\left[2.5,5.5\right] ~\mr{mHz}$.\\

For each model and frequency, between $150$ and $1500$ packets have been pseudo-randomly initialised near the stellar surface, with equal probability per unit area. As pointed out by \cite{Osaki1990}, although almost all of the acoustic emission is expected to occur at the point where convective velocities reach a maximum (see Fig.\,\ref{fig:convection}), some frequencies in ${\cal F}$ may be non-propagating there. \cite{Osaki1990} shows that this evanescent region is responsible for the low frequency behaviour of the internal cavity excitation rate. In our ``classical" description, wave packets do not cross an evanescent region. We therefore pick, somewhat arbitrarily, the near surface layer $T=10^5~\mr{K}$, where all frequencies in ${\cal F}$ are propagating, to be the effective stellar surface for emission. As discussed in \S\ref{sec:propRays} and \S\ref{sec:propTrapping}, we confined the propagation of the packets to the stellar envelope by setting the angle $\cos(\theta)=\hat{k}_0\cdot\hat{g}$ according to the $\theta(\omega)$ relation suitable for $\ell = 50$ packets in $S1$. Here $\hat{k}_0$ is the direction of a packet's initial wave vector $\vec{k}_0$. The remaining angular degree of freedom was picked isotropically. The path of each packet was then integrated (as discussed in \S\ref{sec:propRays}) until either its maximal lifetime $t_\mr{max}(\omega)$ had elapsed, or until the packet was absorbed, whichever occurred first. 

\subsection{Mean packet lifetimes}

\begin{figure}
\includegraphics[width=0.5\textwidth]{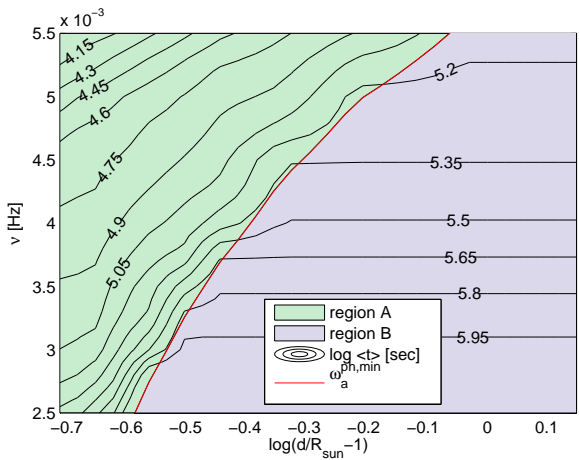}
\caption{Mean packet lifetime as a function of frequency $\nu$ and binary separation parameter $d$. The red curve is the minimal photospheric acoustic cutoff frequency $\omega_a^\mr{ph,min}$ discussed in \S\ref{sec:propRays}.}
\label{fig:meanTime}
\end{figure}

\begin{figure}
\includegraphics[width=0.5\textwidth]{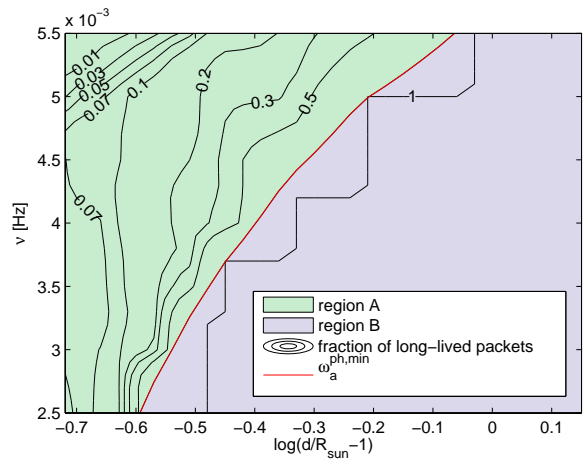}
\caption{Fraction of long-lived packets having frequency $\nu$ propagating in a binary model with separation parameter $d$. The red curve is the minimal photospheric acoustic cutoff frequency $\omega_a^\mr{ph,min}$.}
\label{fig:longFraction}
\end{figure}

The mean packet lifetime $\left<t\right>$, averaged over the ensemble of initial conditions defined above, was computed for each model and frequency. The results are summarized in Fig.\,\ref{fig:meanTime}.
As noted in \S\ref{sec:propTrapping}, all packets having a frequency $\omega$ lower than the minimal photospheric acoustic cutoff $\omega_\mr{a}^\mr{ph,min}$ should be trapped. We call packets reaching their maximal lifetime $t_\mr{max} = 1/2\eta(\omega)$, long-lived packets, so that trapped packets are also long-lived packets. We therefore expect all packets below the $\omega_\mr{a}^\mr{ph,min}(d)$ curve in Fig.\,\ref{fig:meanTime}, to be long-lived. We denote the regions above and below $\omega_\mr{a}^\mr{ph,min}(d)$ as regions A and B respectively. The fraction of long-lived packets for a given frequency and binary separation is shown in Fig.\,\ref{fig:longFraction}. As expected, this fraction reaches unity in region B, with the remaining discrepancy between the $1.0$ contour and $\omega_\mr{a}^\mr{ph,min}(d)$ in Fig.\,\ref{fig:longFraction} being an artefact of the separation and frequency resolutions. As the separation parameter is decreased, the region in which $\omega_\mr{a}^\mr{ph}<\omega$ grows (see Fig.\,\ref{fig:omegaAofTheta}), and more packets are absorbed before reaching maximal lifetime. In Fig.\,\ref{fig:longFraction}, the fraction of long-lived packets is thus reduced for decreasing values of $d$. 

Returning to Fig.\,\ref{fig:meanTime} we see that in region B, the mean packet lifetime does not depend on $d$. This is simply due to the fact that all packets in region B are long-lived. On the other hand, in region A, we find that for a given frequency, the mean lifetimes drop by a factor of between $4$ to $10$ as the separation parameter is decreased to the critical separation of $S13$. Roughly speaking, a reduction in the average lifetime in a given frequency implies that the amount of energy present in this frequency (in steady state), should be correspondingly reduced \citep[see e.g.,][]{Samadi2009}.
The mean lifetime averaged over only the short-lived packets in region A was found to be comparable to the mean lifetime of all packets in this region. This is due to the rapid decline of the fraction of long-lived packets in region A. 

\subsection{Acoustic energy density at the surface}
 
For a specific model and frequency, we compute the relative energy density in a given volume element of the star by measuring the amount of time spent in the element, summed over all packets of the aforementioned ensemble. We choose to do so in the outer portion of the stellar models, between a surface one wavelength below the base of the photosphere and the photosphere's outer edge. We expect this to be closely related to what an observer would see. This density was computed for the same models and frequencies as those of the previous section.

We find that the surface density exhibits two types of behaviour depending on whether the frequency and separation belong to the aforementioned regions A or B.

\begin{figure*}
\centering
\subfloat[All packets]{\label{fig:density0510all}\includegraphics[width=0.337\textwidth]{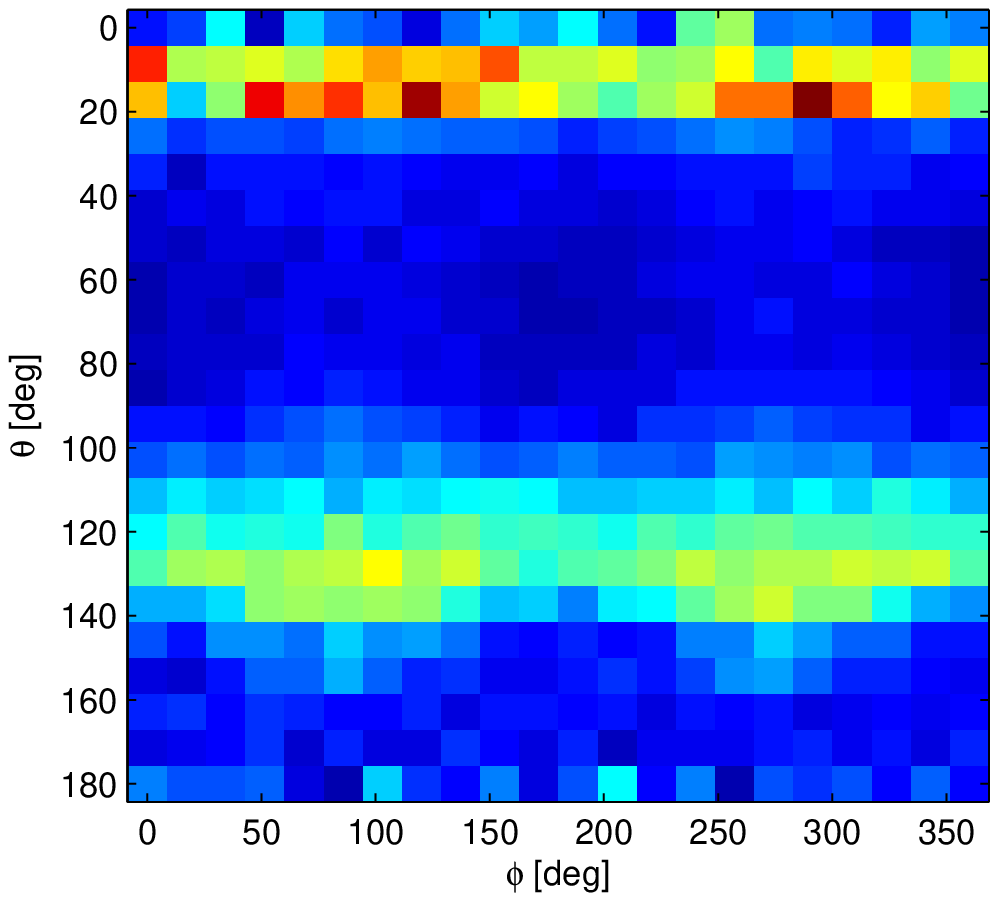}}
\subfloat[Long-lived packets]{\label{fig:density0510long}\includegraphics[width=0.337\textwidth]{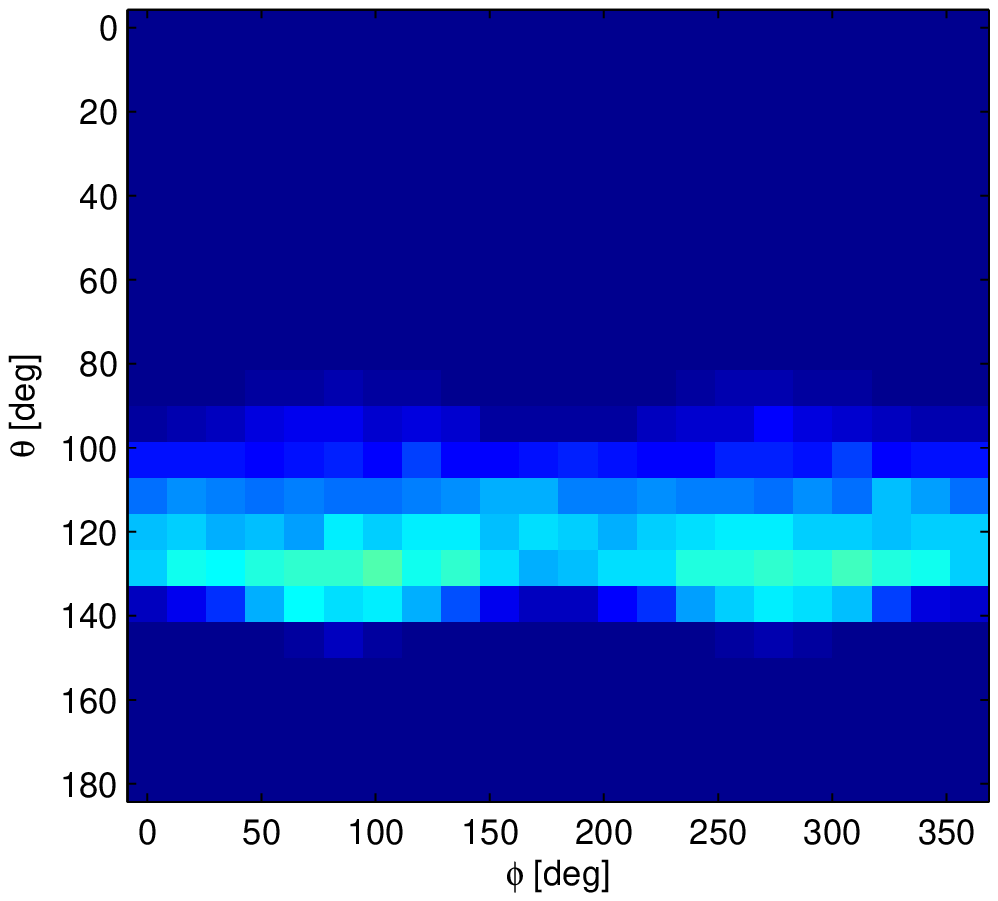}}
\subfloat[Short-lived packets]{\label{fig:density0510short}\includegraphics[width=0.40\textwidth]{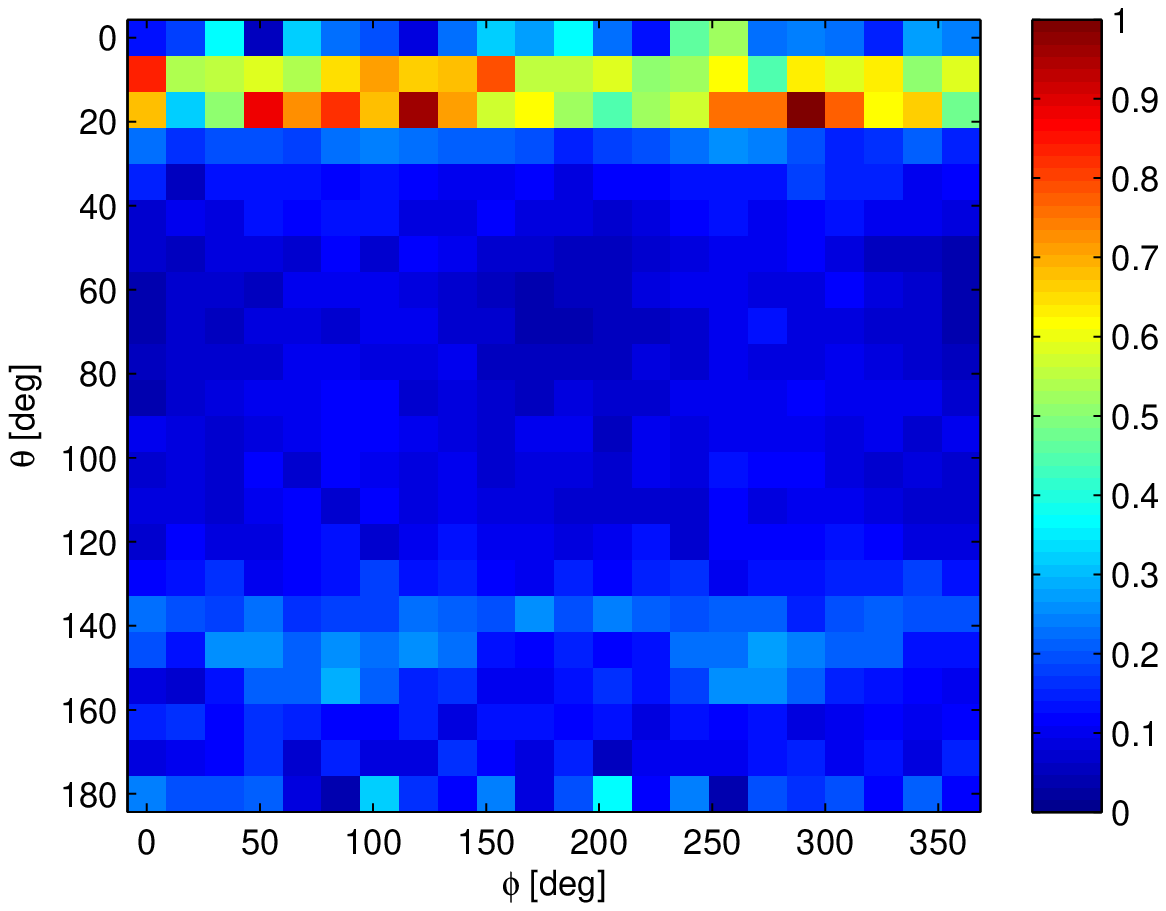}}
\caption{Relative energy density of $3.5 ~\mr{mHz}$ wave packets at the surface of model $S11$ which has a separation parameter $\log(d/R_\odot-1)\approx-0.66$. This result is typical in region A (see text). The angles $\theta$ and $\phi$ are as defined in Fig.\,\ref{fig:titleWithAxes}.}
\label{fig:density0510}
\end{figure*}

In region A, both short-lived packets and long-lived packets coexist. In Fig.\,\ref{fig:density0510} we show the total energy density, and separately, the energy densities contributed by long-lived and short-lived packets. 
Here we see that the energy density of long-lived packets is localized in a band for which $\theta$ is roughly in the range ${\cal B}\equiv\left[110^\circ,140^\circ\right]$. We also see that there is virtually no long-lived contribution outside this band. This is probably because any packet which does not have a path confined to this band, will eventually reach a region where it is absorbed (either near the $L_1$ region or, less likely, at the opposite side; see Fig.\,\ref{fig:omegaAofTheta}). Because these long-lived packets are trapped in this band, and reach their maximal lifetime, we expect them to interfere with themselves and form standing modes in this region.  

The energy density contributed by short-lived waves is just as interesting. In Fig.\,\ref{fig:density0510} we see that this density peaks near the $L_1$ point, where most of these waves are absorbed. There is a small enhancement of this energy density near the band ${\cal B}$. Since it cannot be due to absorption of waves (because in this region $\omega_{\mr{a}}^\mr{ph}$ is greater than $3.5 ~\mr{mHz}$), it is probably a contribution from paths which spent a large fraction of their time near this band, but eventually propagated away.

Referring to Fig.\,\ref{fig:longFraction}, we see that the fraction of long-lived packets corresponding to the frequency and separation parameter of Fig.\,\ref{fig:density0510} is smaller than $0.1$. Although this fraction is small, we also find that the mean lifetime of long-lived packets is larger than the mean lifetime of short-lived packets in this case by approximately one order of magnitude. Evidently, the contribution of each of these populations to the total density is comparable.
In region B, as we have already noted, all packets are long-lived. The density then depends on the relative time spent by this single population at the different surface regions. As the packets propagate throughout a star, they spend a large fraction of their time at the uppermost layers in which the sound speed is small (cf., the sound speed profile of $S1$ is shown in Fig.\,\ref{fig:soundSpeed}). In binary models, surface regions in which $g$ is smaller have a diminished stratification, and therefore a thicker layer of low sound speed. As a result the energy density in these regions is accordingly enhanced. This effect is most significant near the $L_1$ point in the more deformed models. In Fig.\,\ref{fig:density03all}, we see some typical energy densities in region B and their variation as a function of the separation parameter $d$. With increasing values of $d$, the model becomes less deformed, such that both $g$ and the energy density become more homogeneous. 
\begin{figure}
\includegraphics[width=0.5\textwidth]{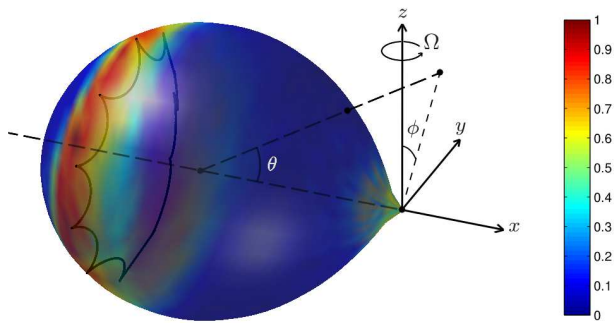}
\caption{Relative total energy density of $3.5 ~\mr{mHz}$ wave packets at the surface of the Roche lobe filling model $S13$ with the path of a specific packet trapped in the band superimposed. The coordinate system is such that the binary system's centre of mass is at the origin, its angular momentum is in the positive $z$ direction, and the stellar core is on the negative $x$ axis.}
\label{fig:titleWithAxes}
\end{figure}
\begin{figure*}
\centering
\subfloat[$\log(d/R_\odot-1)\approx-0.46$]{\label{fig:density0306all}\includegraphics[width=0.337\textwidth]{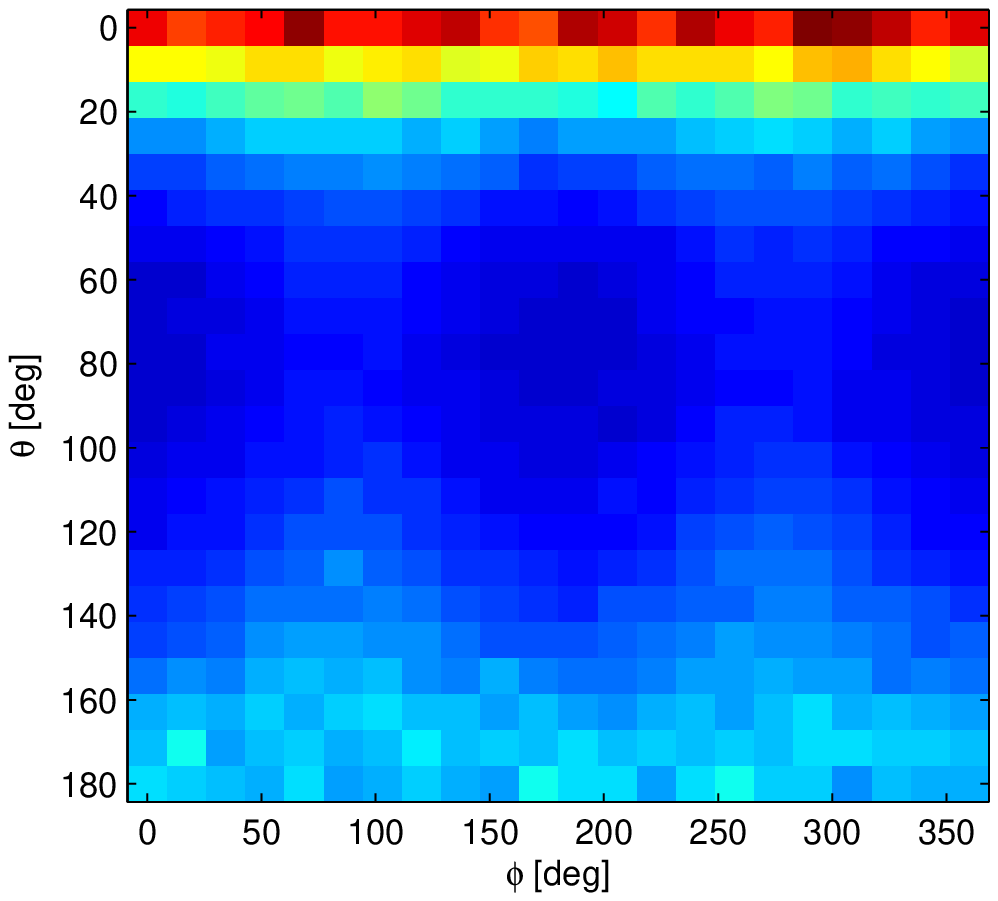}}
\subfloat[$\log(d/R_\odot-1)\approx-0.35$]{\label{fig:density0305all}\includegraphics[width=0.337\textwidth]{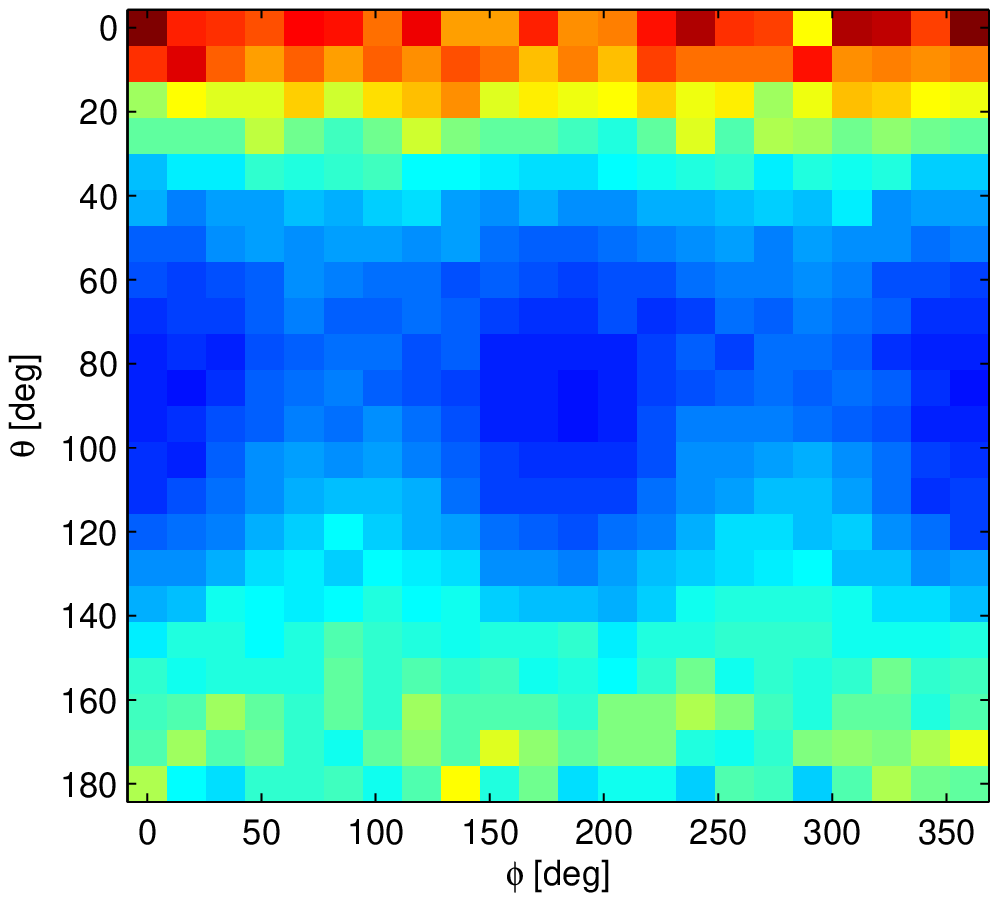}}
\subfloat[$\log(d/R_\odot-1)\approx 0.08$]{\label{fig:density0302all}\includegraphics[width=0.400\textwidth]{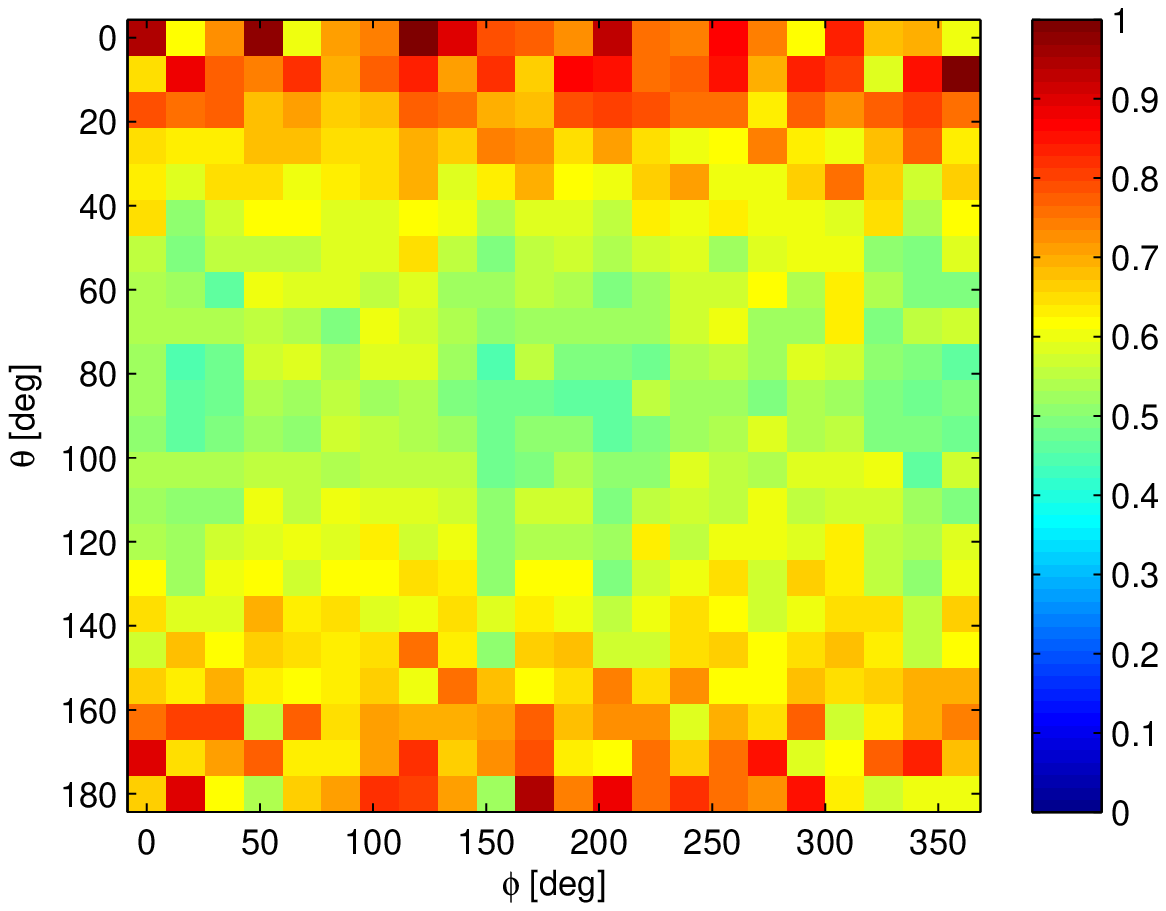}}
\caption{Relative total energy density of $3 ~\mr{mHz}$ wave packets at the surface of models $S7$, $S6$, and $S3$ respectively. Such a dependance on the binary separation parameter $d$ is typical in region B (see text). The angles $\theta$ and $\phi$ are as defined in Fig.\,\ref{fig:titleWithAxes}.}
\label{fig:density03all}
\end{figure*}

\begin{figure}
\includegraphics[width=0.4\textwidth]{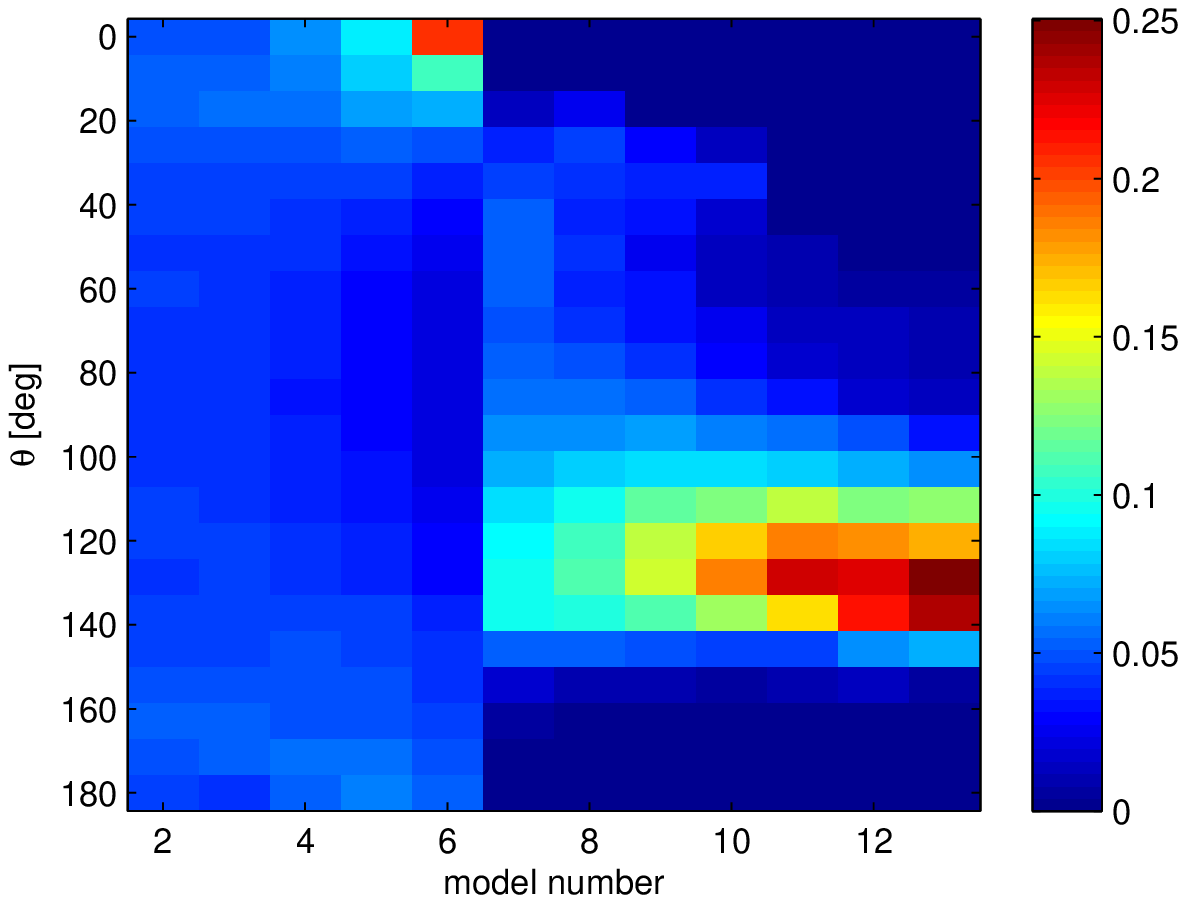}
\caption{Relative energy density averaged over angle $\phi$ at frequency $4.5 ~\mr{mHz}$ as a function of angle $\theta$ and model number. The columns were normalized to unity. The angle $\theta$ here is the one defined in Fig.\,\ref{fig:titleWithAxes}.}
\label{fig:densityStrips450}
\end{figure}

In Fig.\,\ref{fig:densityStrips450} we plot the energy density of long-lived packets averaged over angle $\phi$ as a function of the model number at a fixed frequency. This figure demonstrates a transition between the two types of behaviour previously observed in regions A and B. For the more spherical models (small model numbers) there is no absorption. Waves are trapped in the whole star in this case. Above a critical deformation of the star (at a separation parameter $d$ crossing the curve $\omega_\mr{a}^\mr{ph,min}$ in Fig.\,\ref{fig:longFraction}; this occurs near model $S6$; see also Table \ref{tbl:modelResultingParameters}), a large fraction of the packets end up being absorbed and are therefore quickly drained from the star, leaving only the packets that are trapped in the band ${\cal B}$.


\section{Discussion \& summary}
\label{sec:discussion}

In this work, we have studied the propagation of acoustic waves in close binary stars. Towards this goal, we first developed an approximate stellar numerical model using the KT method, whereby the stellar structure is solved on equipotential surfaces instead of on spherical surfaces. In this method, geometrical correction factors are introduced into the standard set of stellar structure equations. 

We continued by studying the dispersion relation of acoustic waves in the stellar envelope. The propagation of these waves was then solved using a Hamiltonian description of wave packet motion. This allowed us to study the propagation and trapping of waves in the non-spherical models. 

For acoustic waves to be trapped, reflection must occur. Such a reflection can take place near the surface where the wavelength becomes comparable to the local scale height (e.g., the density scale height). This defines an acoustic cutoff frequency below which reflection occurs. For sun-like stars, this reflection gives rise to a range of modes at discrete frequencies. In the non-spherical case, this cutoff frequency becomes location dependent. This implies that waves which should be reflected at one region, may propagate into the upper atmosphere in another. In particular, because the cutoff frequency approaches zero near the $L_1$ point, as the binary fills its Roche lobe, such deformed stars will act as resonant cavities with a ``hole".

Thus, the main result of our work is that a large fraction of waves which otherwise would have been trapped, are actually absorbed near the $L_1$ point. This has important ramifications regarding wave lifetimes and the acoustic energy density at the stellar surface. 

Because the sound crossing time-scale is much shorter than the typical damping time-scales (in spherical stars), waves that can leak out of the cavity are effectively damped by this process. However, it was found that not all waves with a given frequency can reach the $L_1$ point. This gives rise to two populations of waves---short-lived and long-lived waves. 

Because the long-lived waves survive for durations significantly larger than the sound crossing time of the star, they may form modes through constructive interference. Short-lived waves cannot do so. We therefore expect a spectrum which will be composed of both ``narrow" discrete modes due to the long-lived waves and a continuum from the short-lived waves. The ratio between the long-lived and short-lived waves decreases with frequency and as the binary separation approaches the critical value.  

Another very interesting aspect of the long-lived waves, is that in the near Roche lobe filling stars, the region in which they may survive and form modes is a band located at an opening angle of roughly 130$^\circ$ from the $L_1$ point (see Figs.\,\ref{fig:density0510}, \ref{fig:titleWithAxes}, and \ref{fig:densityStrips450}).

Since the visibility of this band should depend on the line-of-sight direction, asteroseismological observations of the star should considerably vary with the binary orbit. This is expected to be a unique feature of such systems. Properties of these modes could be used to place interesting limits on the structure of a binary member, and possibly on the orbital properties of the binary system as well (e.g., the dependence of the spectrum on the binary separation parameter). 

It should be noted that because of the large time-scale disparity (between long-lived and short-lived waves), our conclusions are not very sensitive to the damping mechanism of the trapped waves. Our conclusions are also insensitive to the spectrum of the excitation mechanism. For this reason, we have avoided assuming any particular spectrum. 

The results presented thus far relied on several implicit and explicit assumptions. We note some of these caveats here.

\begin{itemize}
\item Our numerical model assumed that the star is everywhere static. However, we know from von Zeipel's theorem \citep[e.g.,][]{Tassoul1978} that a varying gravitational acceleration necessarily excites motion if the star is to satisfy both the radiative transfer and hydrostatic equations. In rotating stars, this gives rise to the meridional circulation. Here, we should expect a similar flow near the $L_1$ point, but we have neglected it.
\item When discussing the reflection of waves in \S4.3 we have assumed that those waves which are not reflected by the photosphere, but can propagate into it, should be damped quickly and effectively be absorbed there. However, if the star has a corona like the sun, there could in principle be another reflection layer at the transition region between the chromosphere and the corona \citep[cf.][]{BalmforthGough1990}. This transition region can have a sharp density decrease and a sharp temperature increase with an increased acoustic cutoff frequency, about twice as high as in the photosphere. This should allow for higher frequency waves to be trapped in the star. Nevertheless, because the reduction in the acoustic cutoff frequency near the $L_1$ point can be arbitrarily large, our qualitative conclusions remain valid, since a Roche lobe filling star will always have waves which can leak from the stellar acoustic cavity. Namely, for a given frequency we expect the same damping rates we have calculated to occur at somewhat smaller binary separations.
\item In the present work, we concentrated on waves which correspond to modes having an $\ell$-degree of $50$ in spherical stars, but this should be generalized. 
\end{itemize}

Given that the present work is the first analysis of acoustic wave motion in binary members, we could not address many issues which should be dealt with in future work. Following are several examples:

\begin{itemize}
\item One clear extension of our work would be to solve for the acoustic waves in the more realistic non-ZAMS binaries, such as in binary giants. The latter is particularly interesting because it is generally easier to detect the acoustic oscillations in giants. We also expect a large fraction of Roche lobe filling stars to be giants.
\item We have seen that the deformed stars can sustain trapped modes in a band. It would be very interesting to know what these eigenmodes look like. To link between the observations of such modes and the internal structure of the observed star, being able to compute these eigenmodes is essential. One appropriate method is that of Einstein-Brillouin-Keller \citep[e.g., see][]{Swisdak1999}, with which these eigenmodes can be obtained by extending the presented Hamiltonian formalism for the packets.
\item The stellar models we have developed do not describe an overflowing star, as those contain a flow through the $L_1$ point. It would be interesting to see how such a flow could affect the reduced trapping mechanism which we have presented.
\item In this work, we have avoided a discussion of the wave excitation mechanism. We did so for several reasons. First, since there is still no consensus regarding the mechanism responsible for the excitation, we saw no reason to commit to a particular spectrum. Secondly, our main conclusions in this work do not depend on the exact excitation spectrum. Nevertheless, if we wish to calculate the actual binary acoustic spectrum that would be observed, modelling the excitation spectrum is necessary. 
\end{itemize}


\section{Acknowledgements}
\label{sec:acknowledgements}
The authors would like to thank the anonymous reviewer for his valuable comments and suggestions.


\appendix
\section{Binary stellar structure}
\label{sec:structure}
As discussed in this work, the non-spherical structure of close binary stars is paramount to their acoustic behaviour. We thus developed a stellar structure code for synchronously rotating binary stellar models. This code is succinctly described in the following three parts. In \S\ref{sec:structureGeneral} we describe the generic aspects of the code, unrelated to the non-sphericity. To simplify the computation, we focus on zero-age main-sequence stars in which we take the chemical composition to be homogeneous. To enable comparison with existing results in helioseismology, we also confine ourselves to sun-like stars, that is, to $1\mr{M}_\odot$ models having solar composition. In \S\ref{sec:kt} we first summarize the Kippenhahn-Thomas method (KT). Approximating the shape of the binary equipotentials in a co-rotating coordinate system using the binary Roche model, we then apply the KT method to solve for the structure of binary members. In \S\ref{sec:resultingStellarModels}, we describe the resulting models denoted $S1$-$S13$. 

\subsection{General aspects of the stellar structure code}
\label{sec:structureGeneral}

\subsubsection{Internal structure and boundary conditions}
\label{sec:internalStructure}

Assuming thermal and mechanical steady state, we write the well known spherical stellar structure equations 
\begin{subequations}
\label{eq:stellarStructureIndepM}
\begin{align}
\frac{dP}{dM} &= -\frac{G M}{4\pi r^4}\label{eq:dPdM},\\
\frac{dr}{dM} &= \frac{1}{4\pi r^2 \rho}\label{eq:drdM},\\
\frac{dL}{dM} &= \epsilon\label{eq:dLdM},\\
\frac{dT}{dM} &= -\frac{3\bar{\kappa} L_\mr{r}}{64\pi^2 acT^3 r^4},\label{eq:dTdM}
\end{align}\end{subequations}
for the pressure $P$, radius $r$, total luminosity $L$ and temperature $T$ of a stellar layer \citep[see][]{CoxGiuli1968}. Here we use the mass $M$ contained below the stellar layer $r$ as the independent coordinate. This choice is practical when one wants to find stellar models having a set mass as we do. According to the Schwarzschild instability criterion, the radiative luminosity $L_\mr{r}$ is defined as
\begin{equation}
\label{eq:radiativeLuminosity}
L_\mr{r} \equiv
 \begin{cases}
 L & \nabla_\mr{r} - \nabla_\mr{ad} < 0 \\
 L - L_\mr{con} & \text{otherwise},
 \end{cases}
\end{equation}
in the radiative and convective stellar layers respectively. Here we used the radiative gradient
\begin{equation}\label{eq:gradientRadiative}
\nabla_\mr{r} \equiv 
\left(\frac{d\ln T}{d\ln P} \right)_\mr{r} =
\frac{3\bar{\kappa}P L}{16\pi acT^4 GM},
\end{equation}
to express the temperature-pressure stratification in a radiative layer. To model the convective luminosity $L_\mr{con}$, we adopted the version of MLT presented in \citet{KippenhahnWeigert1994}. The equation of state $\rho(T,P)$, Rosseland mean mass opacity $\bar{\kappa}(T,\rho)$, and nuclear energy generation rate $\epsilon(T,\rho)$, appearing in eqs.\ \ref{eq:stellarStructureIndepM}, in addition to the thermodynamic quantities $c_P$, $\delta$ and $\nabla_\mr{ad}$, which are needed to compute $L_\mr{con}$ and to evaluate the Schwarzschild criterion, are all defined in \S\ref{sec:stateFunctions}.

Once these state functions are given, and setting four boundary conditions, the set of equations (\ref{eq:stellarStructureIndepM}) for the quantities $P$, $r$, $L$ and $T$, may be integrated, revealing the stellar structure. Two natural boundary conditions can immediately be stated at the coordinate centre, $M=0$. These are $L_\mr{c} \equiv L(0) = 0$ and $r_\mr{c} \equiv r(0) = 0$. At the outer boundary of the star, we make a distinction between the conditions at the base of the photosphere and those at the stellar surface. We define the base of the photosphere as the layer in which the effective temperature $T_\mr{e} \equiv (L_\mr{s}/4\pi r_\mr{s}^2\sigma)^{1/4}$ is equal to the actual temperature $T$, where $\sigma = ac/4$ is the Stefan-Boltzmann constant (in the following we will also use the subscript $\mr{``e"}$ to denote base-photospheric values of other quantities). The surface is defined to be a layer in which $P_\mr{g} \ll P$. Both at the base of the photosphere and at the surface, zero boundary values for $T$ and $P$ do not strictly hold. The structure of stars having convective envelopes is very sensitive to their actual surface values \citep[cf.,][Ch.\,6]{Clayton1984}. We therefore write two non-zero boundary conditions at the base of the photosphere 
\begin{subequations}
\begin{align}
P_\mr{e} &= P(T_\mr{e}),\\
T_\mr{e} &= (L_\mr{s}/4\pi r_\mr{s}^2\sigma)^{1/4},
\end{align}\end{subequations}
where $P_\mr{e}$ is obtained by separately solving the photospheric structure problem. The fact that $L_\mr{c}$, $r_\mr{c}$, $P_\mr{e}$ and $T_\mr{e}$ are provided at the two opposite integration boundaries presents a practical difficulty. We deal with this by first guessing $L_\mr{e}$, $r_\mr{e}$, $P_\mr{c}$ and $T_\mr{c}$, and then iteratively modifying them until inward and outward integrations match at some intermediate point \citep[see][Ch.\,21]{CoxGiuli1968}.

\subsubsection{Structure of the photosphere}\label{sec:photosphere}

We assume that energy transport in the photosphere is radiative, and that to a very good approximation, $L\approx L_\mr{s}$ and $r\approx r_\mr{s}$. We therefore locally neglect the thin photosphere's curvature and treat it as a plane-parallel slab in which $F \equiv L/4\pi r^2 =\mr{const}$.
Assuming also that the source function is isotropic, the following relation between radiation pressure and flux can be derived from the equations of steady-state radiative transfer \citep[see e.g.][Ch.\,4]{BowersDeeming1984} 
\begin{equation}\label{eq:radiationPressureChange}
\frac{dP_\mr{r}}{dr} = -\frac{\kappa\rho}{c} F.
\end{equation}
This holds exactly for a grey atmosphere in which $\kappa=\mr{const}$, and approximately for non-grey atmospheres when setting $\kappa$ to the Rosseland mean $\bar{\kappa}$, as we do. We define the Rosseland mean optical depth in the radial direction, $-d\tau \equiv \bar{\kappa}\rho dr$ and integrate Eq.\,(\ref{eq:radiationPressureChange}) to obtain
\begin{equation}\label{eq:radiationPressureIntegrated}
P_\mr{r}(\tau) = \frac{F}{c}\tau + P_\mr{r}(0).
\end{equation}
Employing the Eddington approximation and assuming that the radiation intensity is locally Planckian at a temperature $T$ one obtains the following $T(\tau)$ dependence in the photosphere
\begin{equation}
T^4 = \frac{3}{4}T_\mr{e}^4\left(\tau + 2/3\right).
\end{equation}
In this model photosphere, the base is at a mean optical depth of $\tau=2/3$, and at the surface, a temperature minimum of $T(0)=(1/2)^\frac{1}{4}~T_\mr{e} \approx 0.841~T_\mr{e}$ is reached. To find $P_\mr{e}$, the pressure at the base of the photosphere, we integrate the hydrostatic equation 
\begin{subequations}\label{eq:photosphericPressure}
\begin{align}
dP &= \frac{g}{\bar{\kappa}} d\tau,\\
P(\tau) &= P_\mr{r}(0) + \int_0^{\tau} \frac{g}{\bar{\kappa}(\tau')} ~d\tau',\label{eq:photosphericPressureIntegral}
\end{align}\end{subequations}
obtaining $P_\mr{e} = P(T=T_\mr{e}) = P(\tau=2/3)$. In (\ref{eq:photosphericPressureIntegral}) we have assumed that the gas pressure vanishes at the surface. Once $P(T)$ is known, the radius and the mass coordinate of the photospheric layers are calculated by separately integrating ($\ref{eq:dPdM}$) and ($\ref{eq:drdM}$) from the base of the photosphere up to the surface. 

\subsubsection{State functions}\label{sec:stateFunctions}

\begin{figure}
\includegraphics[width=0.5\textwidth]{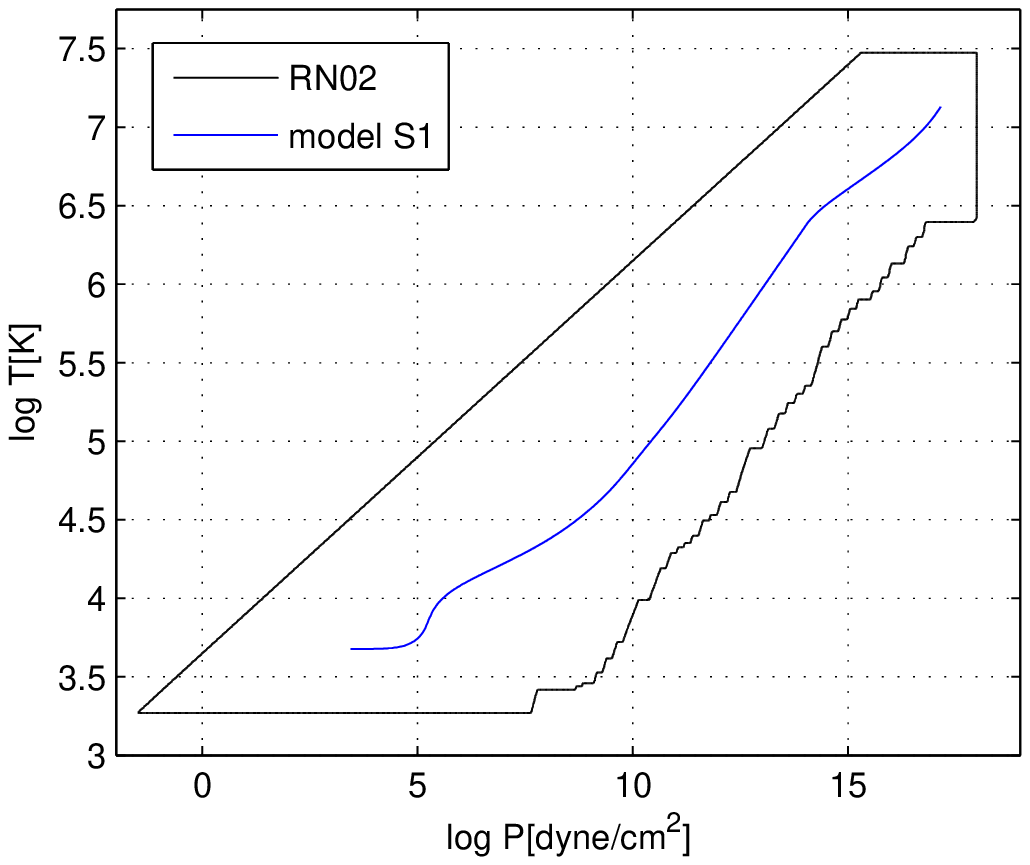}
\caption{Pressure-temperature region in which the equation of state tables (RN02) are defined, containing the $T(P)$ profile of stellar model $S1$.}
\label{fig:eos}
\end{figure}

\begin{figure}
\includegraphics[width=0.5\textwidth]{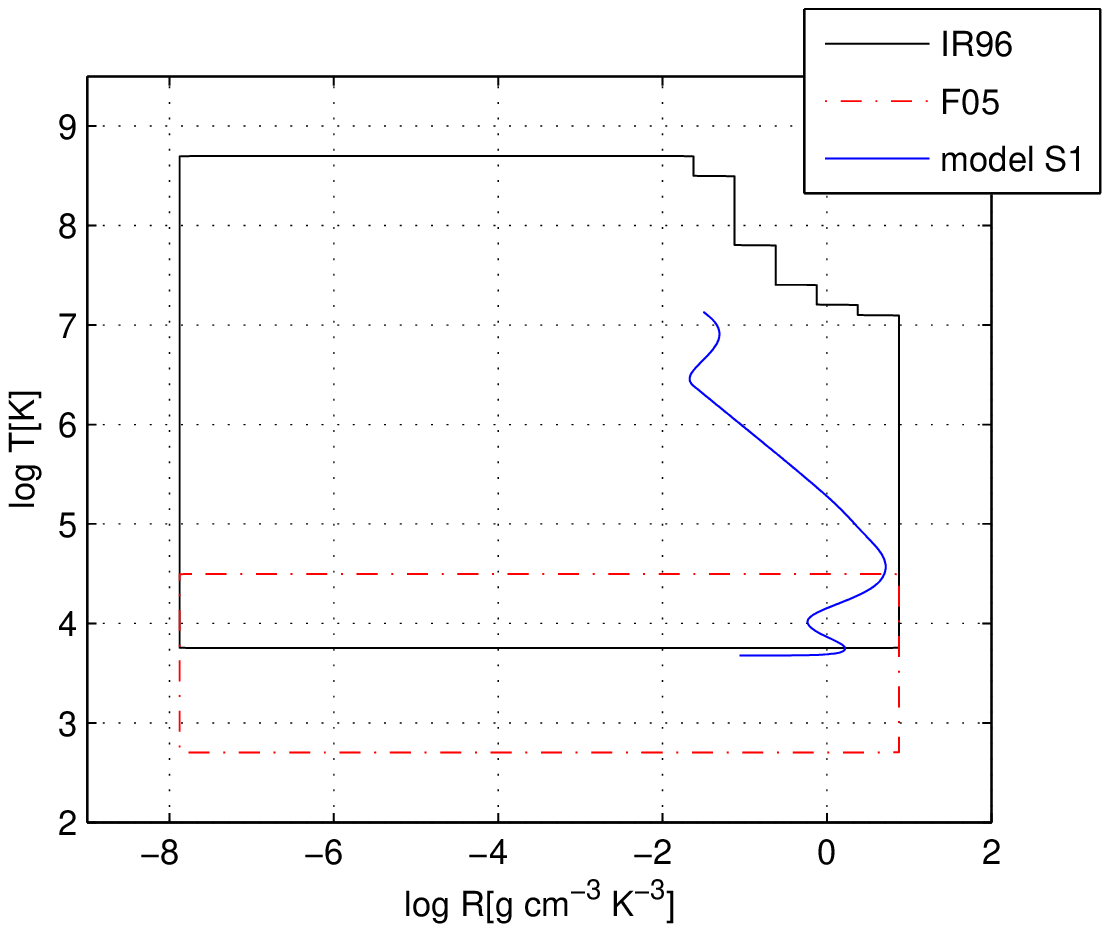}
\caption{$R$-$T$ regions in which the opacity tables (IR96 and F05) are defined, containing the $R$-$T$ profile of stellar model $S1$.}
\label{fig:opacity}
\end{figure}

According to a reference zero-age solar model (denoted as $S0$ and described in \S\ref{sec:resultingStellarModels}), we set the mass fraction of hydrogen to $X=0.70$, and the metallically to $Z=0.02$. For the Rosseland mean opacity and equation of state, tables provided by the OPAL project have been used. The opacity code is described in \citet{IglesiasRogers1996} (IR96), and the equation of state code is described in \citet{RogersNayfonov2002} (RN02). The RN02 tables provide the following quantities as a function of total pressure $P$ and temperature $T$: the density $\rho$, the first and second adiabatic exponents $\Gamma_1$ and $\Gamma_2$, the specific heat at constant density $c_\rho$, and the quantities $\chi_\rho$ and $\chi_T$. These are defined as follows
\begin{align*}
\Gamma_1 \equiv \left(\frac{\dpa\ln P}{\dpa\ln\rho}\right)_\mr{ad},~~~~
\frac{\Gamma_2}{\Gamma_2-1} \equiv \left(\frac{\dpa\ln P}{\dpa\ln T}\right)_\mr{ad},\\
c_\rho \equiv \left(\frac{\dpa E_\mr{i}}{\dpa T}\right)_\rho,~~~~
\chi_\rho \equiv \left(\frac{\dpa\ln P}{\dpa\ln \rho}\right)_T,~~~~
\chi_T \equiv \left(\frac{\dpa\ln P}{\dpa\ln T}\right)_\rho.
\end{align*}
From the above quantities, the following were derived \citep[see][Ch.\,9]{CoxGiuli1968}
\begin{align*}
\nabla_\mr{ad} = \frac{\Gamma_2-1}{\Gamma_2},~~~~
\delta = \frac{\chi_T}{\chi_\rho},~~~~
c_P = c_\rho + \frac{P}{\rho T}\frac{\chi_T^2}{\chi_\rho^2},
\end{align*} 
where $c_P$ is the specific heat at constant pressure. In Fig.\,\ref{fig:eos} we show the region of $P$-$T$ space in which the tables of RN02 are defined, and a $T(P)$ profile of a spherical zero-age $1 \mr{M}_\odot$ model produced by our stellar structure code (this is model $S1$ of \S\ref{sec:resultingStellarModels}).
For the photospheric calculations, described in \S\ref{sec:photosphere}, the values of the opacity at temperatures $\log (T/\mr{K}) < 3.75$ were required, but were not available in IR96. These were provided by the tables of \citet{FergusonAlexander2005} (F05). The tables of IR96 and F05 specify the mean opacity $\bar{\kappa}$, as a function of the temperature $T$, and $R\equiv\rho_\mr{cgs}/(T_6)^3$, where $T_6\equiv 10^{-6}\times (T/\mr{K})$. In the overlap region between IR96 and F05 (see Fig.\,\ref{fig:opacity}), the opacity values of IR96 were used.
All of the above tables were provided for various compositions ($X$ and $Z$), using the metal abundance ratios of the present-day sun, as found by \citet{GrevesseNoels1993}. From these raw tables, composition specific tables were interpolated using code provided by the OPAL project. 

\begin{figure}
\includegraphics[width=0.5\textwidth]{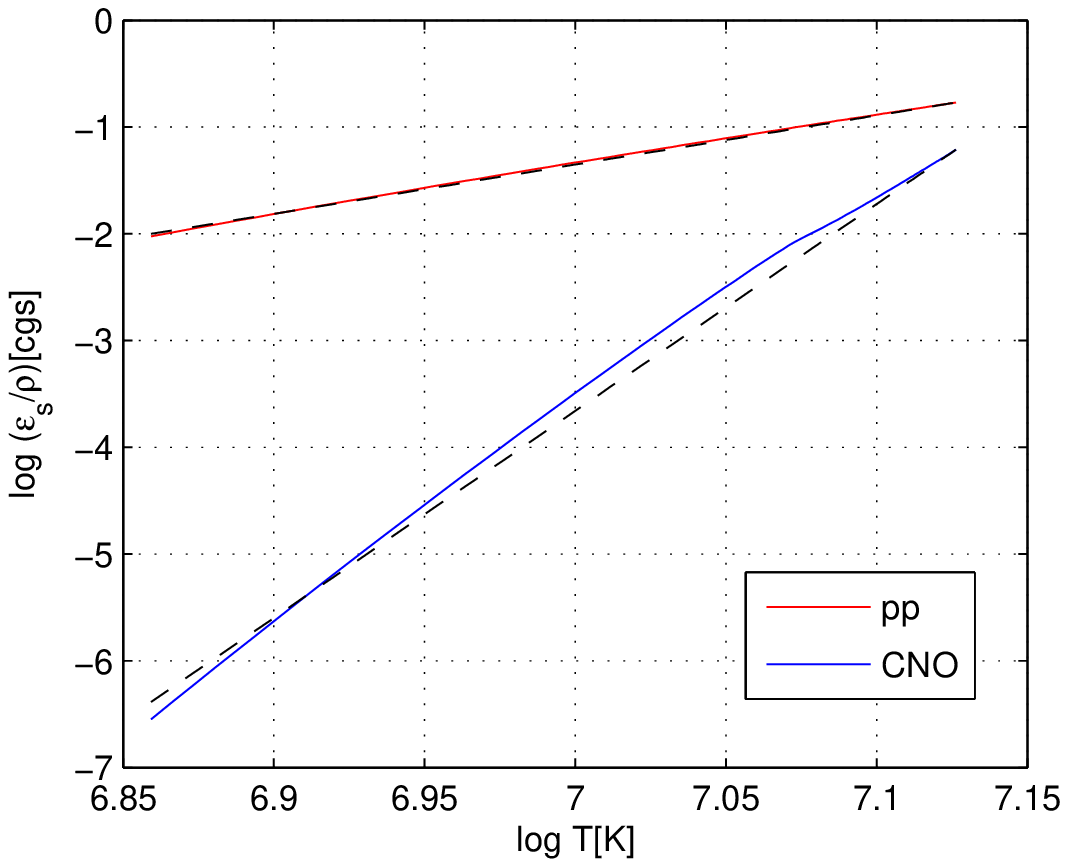}
\caption{Linear fit of $\log \epsilon_\mr{pp}/\rho$ and $\log \epsilon_\mr{CNO}/\rho$ as functions of $\log T$, in model $S0$.}
\label{fig:nuclear}
\end{figure}

We now proceed to define the specific nuclear energy generation rate, $\epsilon(T,\rho)$. Two series of nuclear reactions account for almost all of the non-neutrinic solar luminosity. These are the proton-proton chain (p-p chain) and the CNO cycle. Due to a large temperature sensitivity of these processes (for hydrogen burning $\epsilon_\mr{pp}\sim T^{4.5}$ and $\epsilon_\mr{CNO}\sim T^{18}$ should approximately hold; see e.g. \citealt{BowersDeeming1984}) one can show that almost all of the luminosity in a model of a zero-age sun originates in a relatively narrow temperature range. In reference model $S0$, the fraction of the luminosity produced between the centre at $\log(T/K)\approx 7.13$ and the layer $\log(T/K)\approx 6.86$ was $0.99$ of the surface luminosity. Assuming that both in the p-p chain and in the CNO cycle, a single reaction in the series sets the time-scale for the whole series to operate, then the energy generation rate of each process can be approximated as \citep[see e.g.,][Ch.\,17]{CoxGiuli1968}
\begin{equation}\label{eq:epsilonInterp}
\epsilon_\mr{s} \approx \epsilon_{\mr{s},0}~\rho\left(\frac{T}{T_0}\right)^{\nu_s},
\end{equation}
which should be valid in a small enough temperature range around $T_0$. Here the subscript $\mr{``s"}$ denotes the series (either p-p or CNO), and $\epsilon_{\mr{s},0}=\epsilon_\mr{s}(T_0,\rho_0) / \rho_0$. For the central temperature value of model $S0$, $T_0 = 1.337 \times 10^7 ~\mr{K}$, we obtain the $\nu_\mr{pp}$, and $\nu_\mr{CNO}$ from a $\log$-$\log$ fit of (\ref{eq:epsilonInterp}) to $\epsilon_\mr{pp}(T)/\rho(T)$ and $\epsilon_\mr{CNO}(T)/\rho(T)$ of model $S0$, performed in the temperature range mentioned above. We show this fit in Fig.\,\ref{fig:nuclear}, and the fit parameters in table \ref{tbl:epsilonInterpFit}.
By using the values $\epsilon_{\mr{pp},0}$ and $\epsilon_{\mr{CNO},0}$ of model $S0$, we are effectively using the Debye screening factors at $T_0$, $\rho_0$ in model $S0$, at all values of $T$, $\rho$ in our models.\\

\begin{table}\centering
\begin{tabular}{l||c*{1}c}
& $\nu_s$ & $\epsilon_{\mr{s},0} ~\mr{(cgs)}$ \\
\hline
pp      & $4.608$ & $1.701\times 10^{-1}$ \\
CNO     & $19.39$ & $6.145\times 10^{-2}$ \\
\end{tabular}
\caption{Fit parameters of interpolation function (\ref{eq:epsilonInterp}).}
\label{tbl:epsilonInterpFit}
\end{table}

\subsection{The Kippenhahn-Thomas method}\label{sec:kt}

Up to this point, our discussion of stellar structure ignored the possible effects of rotation and tidal forces. The spherical solutions that may be obtained from equations (\ref{eq:stellarStructureIndepM}) usually provide adequate models of slowly rotating single stars, or in some cases, serve as simplified evolutionary models of binaries undergoing mass transfer via Roche lobe overflow. We, on the other hand, are specifically interested in finding how the effects of rotation and tide alter the envelope structure of close binaries, and in turn, in how this altered structure affects the transport of acoustic energy within them. At first sight, it would seem that this would necessarily require us to solve the three dimensional energy and momentum conservation equations. This would be a numerical task very different from that of solving equations (\ref{eq:stellarStructureIndepM}). Using the method of \citet{KippenhahnThomas1970} (KT), we will now see that an approximate solution can be obtained from a slightly modified version of (\ref{eq:stellarStructureIndepM}). 

Assuming that in some co-rotating coordinate system the stellar material has settled into a static configuration, then the force field can be derived from an effective potential $\phi = \phi_\mr{cent} + \phi_\mr{g}$, having centrifugal and gravitational components. Setting $\vec{v}=0$ in the Euler equation, we write the hydrostatic equation in vector form
\begin{equation}\label{eq:ktEuler}
\nabla\phi = -\frac{1}{\rho}\nabla P,
\end{equation}
which necessarily shows that $P$ and $\phi$ are constants on the same surfaces, or that $P=P(\phi)$. Making use of this fact, we write (\ref{eq:ktEuler}) as 
\begin{equation}\label{eq:ktHydrostatic}
\rho = -\frac{dP}{d\phi},
\end{equation}
which also shows that $\rho=\rho(\phi)$. Using the equation of state $P(\rho,T)$, assuming that the composition is a constant on equipotentials and that the material is in LTE, we see that the other thermodynamic quantities (e.g., $T$) are also functions of $\phi$. Some other scalar quantities, such as the effective gravitational acceleration $g$, are not generally constants on equipotentials. Denoting by $S_\phi$ an equipotential surface for some value $\phi = \mr{const}$, we define the mean value of a quantity $f$ on $S_\phi$ as
\begin{equation}
\mean{f} = \frac{1}{S_\phi}\oint_{S_\phi}f ~ds.
\end{equation}
Using the unit vector $\hat{n} \equiv \nabla\phi/\Vert\nabla\phi\Vert$, and denoting differentiation in the direction of $\hat{n}$ as $d/dn\equiv\hat{n}\cdot\nabla$, we can express the local effective gravitational acceleration as $g=d\phi/dn$, and the normal distance between the equipotentials $\phi$ and $\phi+d\phi$, as $dn = d\phi/g$. Like $g$, the differential distance $dn$ may also vary on equipotentials. We now use the mean values $\mean{g}$ and $\mean{g^{-1}}$, to describe geometric properties of the surfaces $S_\phi$. We express the volume $dV_\phi$ enclosed between $\phi$ and $\phi+d\phi$ as
\begin{equation}\label{eq:ktDV}
dV_\phi = \oint_{S_\phi} dn ~ds = d\phi\oint_{S_\phi} \frac{dn}{d\phi} ~ds = S_\phi\mean{g^{-1}} d\phi.
\end{equation}
Using the volume $V_\phi$ enclosed by $S_\phi$, we define $V_\phi\equiv\frac{4\pi}{3}r_{\phi}^3$, where $r_\phi$ is an effective ``volume radius". For further use, we define the following dimensionless quantities 
\begin{align}\label{eq:ktUVW}
u \equiv \frac{S_\phi}{4\pi r_\phi^2},~~~~
v \equiv \mean{g}\frac{r_\phi^2}{GM},~~~~
w \equiv \mean{g^{-1}}\frac{GM}{r_\phi^2},
\end{align}
which all tend to $1$ as the surface $S_\phi$ becomes more spherical. In definition (\ref{eq:ktUVW}), $M$ is the mass contained in the volume $V_\phi$. We now use the above definitions to write stellar structure equations analogous to (\ref{eq:stellarStructureIndepM}), essentially replacing in the derivations of \ref{eq:stellarStructureIndepM} the surfaces $r=\mr{const}$ with equipotentials. Expressing the mass contained in the differential volume (\ref{eq:ktDV}),
\begin{equation}\label{eq:ktdMphi}
dM=\rho(\phi)dV_\phi=4\pi r_{\phi}^2\rho(\phi)dr_\phi,
\end{equation}
we obtain an equation analogous to (\ref{eq:drdM})
\begin{equation}\label{eq:ktdrdM}{
\frac{dr_\phi}{dM}=\frac{1}{4\pi r_{\phi}^2\rho}}~.
\end{equation}
In the hydrostatic equation (\ref{eq:ktHydrostatic}), we derive the following expression for the potential differential 
\begin{align}\label{eq:ktdphi1}
d\phi = \left(\frac{dV_\phi}{d\phi}\right)^{-1} dV_\phi = \left(\frac{dV_\phi}{d\phi}\right)^{-1} \frac{dM}{\rho} = \frac{dM}{\mean{g^{-1}} S_\phi \rho},
\end{align}
where we have used (\ref{eq:ktdMphi}) in the second equality and (\ref{eq:ktDV}) in the third. Using the definitions (\ref{eq:ktUVW}), we rewrite this result using just dimensionless quantities and the structure variables $r_\phi$ and $M$, as 
\begin{align}
d\phi = \frac{GM}{\rho uw4\pi r_{\phi}^4}dM.
\end{align}
Substituting $d\phi$ into (\ref{eq:ktHydrostatic}), we obtain an equation analogous to (\ref{eq:dPdM})
\begin{equation}\label{eq:ktdPdM}{
\frac{dP}{dM} = -\frac{GM}{uw4\pi r_{\phi}^4} = -\frac{GM}{4\pi r_{\phi}^4} f_P}~,
\end{equation}
where in the last equality we have defined $f_P \equiv 1/uw$. The energy generation rate $\epsilon$, being a function of temperature and density, is also constant on equipotentials for constant composition. Assuming thermal steady state, the energy generated per unit time, $\epsilon dM$, in the volume $dV_\phi$, is exactly balanced by an increase in the outgoing luminosity $L$ of the surface $S_\phi$. We thus obtain 
\begin{equation}\label{eq:ktdLdM}{
\frac{dL}{dM} = \epsilon}~,
\end{equation}
analogous to (\ref{eq:dLdM}). To find the temperature stratification, we express the radiative flux in the direction $\hat{n}$, using the diffusion approximation, 
\begin{equation}\label{eq:ktFr}
F_\mr{r} = -\frac{4 acT^3}{3\bar{\kappa}\rho} \frac{dT}{dn} = -\frac{4 acT^3}{3\bar{\kappa}\rho} g\frac{dT}{d\phi}, 
\end{equation}
where the definition of $dn$ was used. For constant composition, $\bar{\kappa}(T,\rho)$ is a constant on equipotentials. We can therefore integrate (\ref{eq:ktFr}) and substitute the expression (\ref{eq:ktdphi1}) for $d\phi$, to obtain the radiative luminosity through the surface $S_\phi$, 
\begin{equation}
L_\mr{r} = 
\oint_{S_\phi} F_\mr{r} ~ds = 
-\frac{4acT^3}{3\bar{\kappa}}S_{\phi}^2\mean{g}\mean{g^{-1}}\frac{dT}{dM},
\end{equation}
which can be written using the dimensionless quantities (\ref{eq:ktUVW}) as
\begin{equation}
L_\mr{r} = 
-\frac{64\pi^2 acT^3 r_\phi^4}{3\bar{\kappa}}u^2vw\frac{dT}{dM}.
\end{equation}
Defining $f_T \equiv 1/u^2vw$, we obtain the equation analogous to (\ref{eq:dTdM})
\begin{equation}\label{eq:ktdTdM}{
\frac{dT}{dM} = -\frac{3\bar{\kappa}L_\mr{r}}{64\pi^2 acT^3 r_\phi^4}f_T}~.
\end{equation}
To summarize, we see that by defining the new quantity $r_\phi$, and redefining the rest of the variables appearing in (\ref{eq:stellarStructureIndepM}) to represent their respective equipotential values, we obtain an equivalent set of equations (\ref{eq:ktdrdM}, \ref{eq:ktdPdM}, \ref{eq:ktdLdM}, \ref{eq:ktdTdM}), with the additional dimensionless correction factors $f_P$ in (\ref{eq:ktdPdM}) and $f_T$ in (\ref{eq:ktdTdM}). In the spherical case $r_\phi = r$, and $f_P = f_T = 1$, so that the original stellar structure equations are recovered. To solve the new set of equations for a binary star, we will now need to find the functions $f_P$ and $f_T$. We will also need to redefine $L_\mr{con}$, the convective contribution to the total luminosity, and correct the photospheric boundary conditions. These last two issues are dealt with in \S\ref{sec:ktCorrections}.

\subsubsection{Approximate binary potential and corresponding KT functions}\label{sec:binaryPotential}

\begin{figure}
\includegraphics[width=0.5\textwidth]{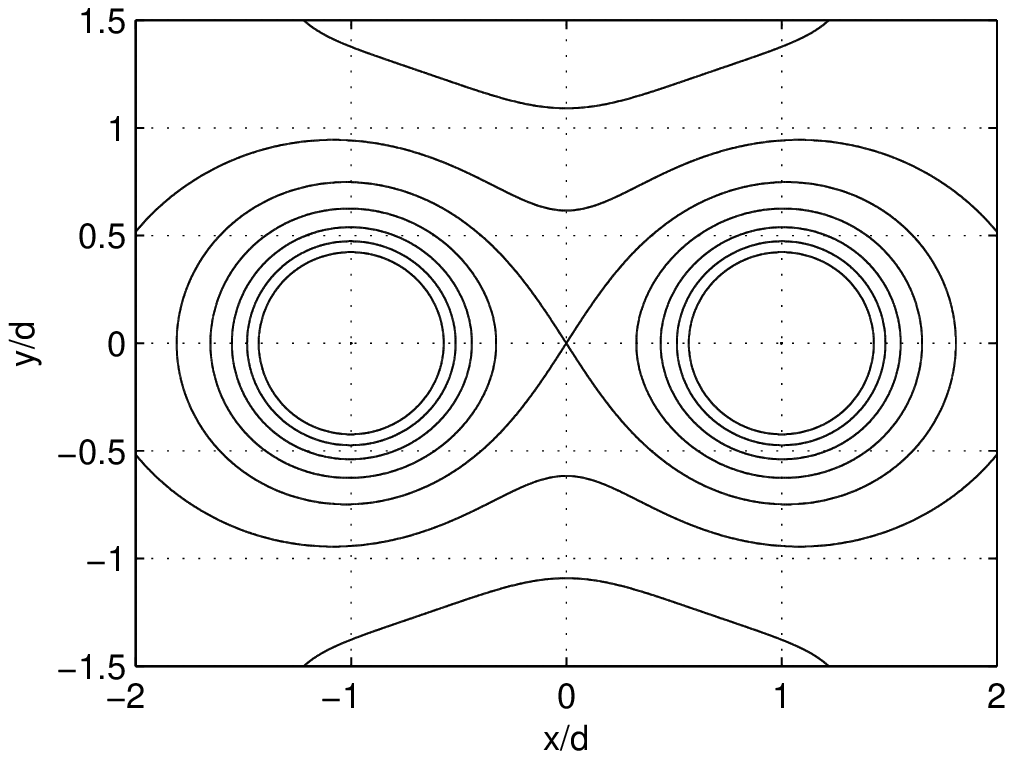}
\caption{Some equipotentials of $\tilde{\phi}$ in the $\tilde{z}=0$ plane, including the critical equipotential $\tilde{\phi}=-2$.}
\label{fig:roche}
\end{figure}

\begin{figure}
\includegraphics[width=0.5\textwidth]{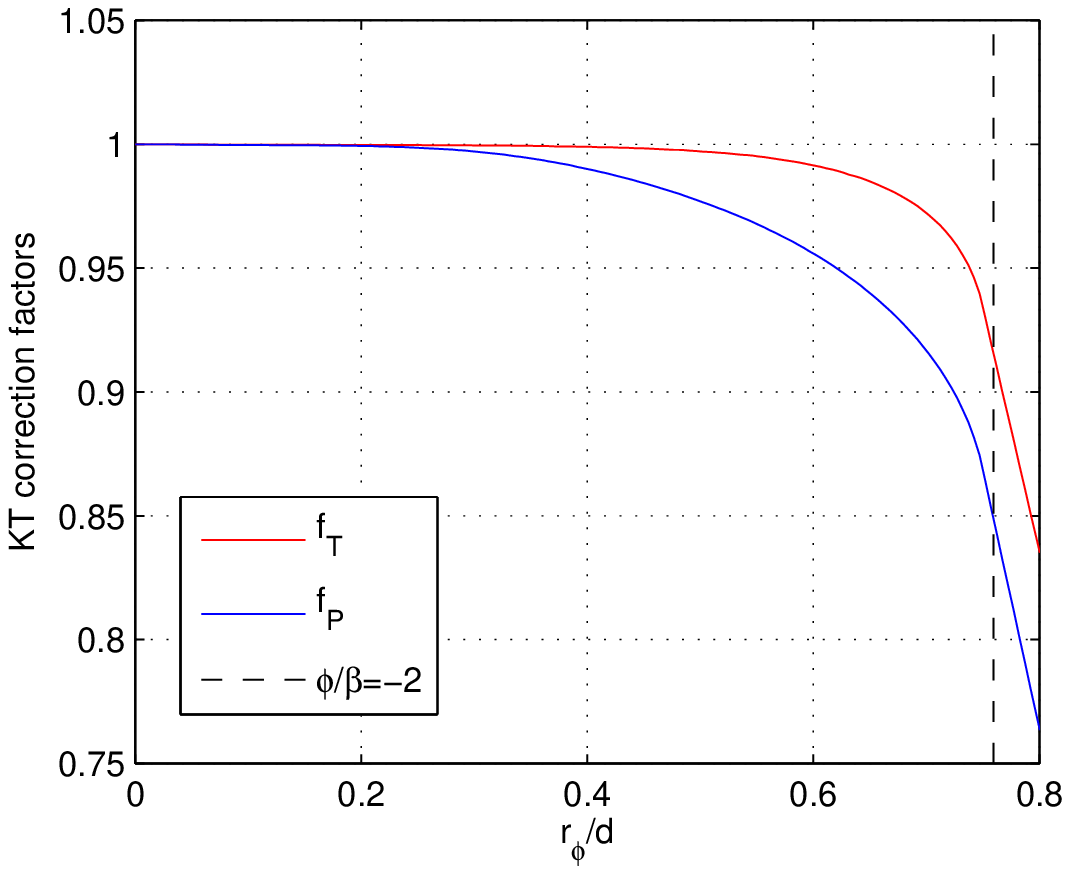}
\caption{KT correction factors for a binary system with a mass ratio of $1$. The values to the right of the dashed line have been linearly extrapolated (see text).}
\label{fig:ktFuncs}
\end{figure}

In the KT formulation for obtaining the stellar structure, we have all the while assumed that the effective potential $\phi$ is known. In a coordinate system co-rotating with the static gas at an angular velocity $\Omega \hat{z}$, with origin at the system's centre of mass, we have 
\begin{equation}\label{eq:ktphi}
\phi(\vec{r}) = -\frac{1}{2}\Omega^2(\vec{r}\times\hat{z})^2 + \phi_\mr{g}(\vec{r}).
\end{equation}
In turn, the gravitational potential $\phi_\mr{g}$ depends on the density through the Poisson equation, 
\begin{equation}\label{eq:poisson}
\nabla^2\phi_\mr{g} = 4\pi G\rho.
\end{equation}
We see then that a full solution of the binary structure problem should simultaneously satisfy (\ref{eq:poisson}) and (\ref{eq:ktdrdM}, \ref{eq:ktdPdM}, \ref{eq:ktdLdM}, \ref{eq:ktdTdM}). In principal, iterative methods could be used. These could start with an initial guess for $\phi$, compute the stellar structure including $\rho$, and then repeatedly adjust $\phi$ until the Poisson equation is sufficiently satisfied. We choose instead to approximate the potential using the Roche model \citep[see][Ch.\,16 and Ch.\,5]{Tassoul1978}. In this model, one assumes, for the purpose of finding $\phi$, that the total mass of each binary member is concentrated at a central point. Due to the fact that in the spherical case, a large fraction of the mass is concentrated at the core, with most of it below the envelope (see \S\ref{sec:resultingStellarModels}), we expect that this approximation should not significantly affect the value of $\phi$ in the envelope. In the central regions of the star, the Roche model's $\phi$ is singular, and so will surely not provide a good approximation there. Fortunately, we only require the geometric properties $f_T$ and $f_P$ of the equipotentials of $\phi$, and not its actual value. Below the envelope, the Roche equipotentials are approximately spheres (see Fig.\,\ref{fig:roche}). We assume this is a reasonable approximation for the shape of the inner equipotentials of an actual binary. Using the Roche model, we express (\ref{eq:ktphi}) as
\begin{equation}\label{eq:ktphi2}
\phi(\vec{r}) = 
-\frac{1}{2}\Omega^2(x^2 + y^2) - 
\frac{GM_1}{\Vert\vec{r}-\vec{r}_1\Vert} - 
\frac{GM_2}{\Vert\vec{r}-\vec{r}_2\Vert},
\end{equation}
where $M_i$ and $\vec{r}_i$ are the mass and co-rotating position of the i'th component. Assuming circular orbits, we obtain a Keplerian frequency $\Omega^2 = G(M_1 + M_2)/D^3$, where $D \equiv \Vert \vec{r}_1 - \vec{r}_2 \Vert$. To simplify the computation of $f_T$ and $f_P$, we confine ourselves to binaries having a mass ratio of 1, so that (\ref{eq:ktphi2}) can be written as
\begin{equation}
\phi(\vec{r}) = 
-\frac{GM}{8d^3}r_{xy}^2 - GM\left(\frac{1}{r_1} + \frac{1}{r_2}\right),
\end{equation}
where we have defined $M = M_i$, $d = D/2$, $r_i = \Vert\vec{r}-\vec{r}_i\Vert$ and $r_{xy}^2 = x^2+y^2$. Defining the dimensionless lengths $\tilde{r}_i = r_i/d$, $\tilde{r}_{xy} = r_{xy}/d$, and a dimensionless potential $\tilde{\phi} = \phi/\beta$, with $\beta = GM/d$, we have that
\begin{equation}
\tilde{\phi} = -\frac{1}{8}\tilde{r}_{xy}^2 - \frac{1}{\tilde{r}_1} - \frac{1}{\tilde{r}_2}.
\end{equation}
Using $\tilde{\phi}$, we can numerically compute the functions $f_P$ and $f_T$, for all combinations of the parameters $M$ and $d$. We now describe how this was done. To simplify the notation we use the variables $\vec{r}$, $\phi$ to denote their dimensionless counterparts $\tilde{\vec{r}}$, $\tilde{\phi}$ (this is equivalent to setting $d = M = G = 1$). Working in the cartesian coordinate system defined above, $\phi(\vec{r})$ and $g(\vec{r})$ were sampled in the positive quadrant of the $x$-$z$ plane ($\phi(\vec{r})$ is symmetric under the operations $z\rightarrow-z$ and $y\rightarrow-y$), at evenly spaced values of $x$, $y$, and $z$. In such a sampling scheme, numerically computing accurate surface integrals required a higher resolution than was required for computing volume integrals. We therefore relate surface and volume integrals in the following way. Denoting by $\Delta V_\phi$ the finite volume between the equipotentials $\phi$ and $\phi+\Delta\phi$, and recalling that $dn=d\phi/g$, we approximately have
\begin{equation}\label{eq:ktVolToSurf1}
\int_{\Delta V_\phi} g^n ~d^3 r \approx 
\oint_{S_\phi} g^n \Delta n ~ds = 
\Delta\phi \oint_{S_\phi} g^{n-1} ~ds,
\end{equation}
which is exact in the limit $\Delta\phi \rightarrow 0$. Rewriting (\ref{eq:ktVolToSurf1}), we obtain
\begin{equation}{
S_{\phi}\mean{g^{n-1}} = \frac{1}{\Delta\phi}\int_{\Delta V_\phi} g^n ~d^3 r}~.
\end{equation}
By setting $n$ to $1$, this can be used to compute the surface area,
\begin{equation}\label{eq:ktVolToSurf3}
S_{\phi} = S_{\phi}\mean{g^0} = \frac{1}{\Delta\phi}\int_{\Delta V_\phi} g ~d^3 r,
\end{equation}
and, once $S_{\phi}$ is known, to find the surface averaged value of $g^n$,
\begin{equation}\label{eq:ktVolToSurf4}
\mean{g^n} = \frac{1}{S_{\phi} \Delta\phi}\int_{\Delta V_\phi} g^{n+1} ~d^3 r.
\end{equation}
Using (\ref{eq:ktVolToSurf3}) and (\ref{eq:ktVolToSurf4}), we computed the quantities $S_\phi$, $V_\phi$, $\mean{g}$, and $\mean{g^{-1}}$ for discrete values of $\phi$, up to the critical equipotential $\phi=-2$. Using definition (\ref{eq:ktUVW}), we found the corresponding values of $u$, $v$, and $w$ (recall that in the Roche model, $M_\phi=\mr{const}=M$, and that we have set the total mass $M$ to $1$), from which the KT functions $f_T$ and $f_P$ were computed as a function of $r_\phi$ (see Fig.\,\ref{fig:ktFuncs}).
Although we will not be computing overflowing models in this work, it was numerically convenient to linearly extrapolate $f_T$ and $f_P$ beyond $\phi=-2$ (corresponding to $r_\phi\approx0.7595$). This enables the stellar structure code to temporarily $``$cross$"$ the critical equipotential, in the process of converging to a sub-critical solution.

\subsubsection{Corrections to the photospheric and convective equations}\label{sec:ktCorrections}

We must now revise the previous treatments of the convection zone (\S\ref{sec:internalStructure}) and the photosphere (\S\ref{sec:photosphere}), to deal with a possible variation of $g$, and $F$ on equipotentials. We start by rewriting the radiative gradient. In a radiative region, we may set $L=L_\mr{r}$ in (\ref{eq:ktdTdM}). Using (\ref{eq:ktdPdM}) we obtain
\begin{equation}\label{eq:ktGradientRadiative}
\nabla_\mr{r} \equiv 
\left(\frac{d\ln T}{d\ln P} \right)_\mr{r} =
\frac{3\bar{\kappa}P L}{16\pi acT^4 GM} \frac{f_T}{f_P},
\end{equation}
with which The Schwarzschild criterion in eqs.\ \ref{eq:stellarStructureIndepM} and \ref{eq:radiativeLuminosity} may now be evaluated.

Throughout most of the convection zone (which spans approximately $0.3$ of the radial extent of $S1$) convection is nearly adiabatic. The details of the MLT treatment become relevant only at a thin layer below the photosphere in which the actual gradient, $\nabla \equiv d\ln T/d\ln P$, becomes significantly larger than $\nabla_\mr{ad}$. For the purpose of calculating the quantities $L_\mr{con}$, and the mean squared convective velocities $\mean{v_\mr{con}^2}$ \citep[appearing in the derivations of MLT in][]{KippenhahnWeigert1994} in this superadiabatic layer, we approximate it as a plane-parallel slab, having a surface area $S_\phi$, in which $g$ and $F$ are assumed constant and equal to $g_\mr{s} \equiv \mean{g}$, and $F_\mr{s} \equiv L/S_\phi$ respectively. This treatment is similar to that of \citet{LandinMendes2009}, where a convective gradient $\nabla_\mr{con}$ is computed using MLT, and taken to be constant on equipotentials. In the photosphere, for the purpose of obtaining the $P(T)$ relation (\ref{eq:photosphericPressure}), we again approximate $g$ and $F$ using the constant values $g_\mr{s}$ and $F_\mr{s}$. Given the surface luminosity $L_\mr{s}$ (see \S\ref{sec:internalStructure}), we redefine the effective temperature as $T_\mr{e} = (L_\mr{s}/S_\phi\sigma)^{1/4}$, and obtain $P_\mr{e} = P(T_\mr{e})$. With the photospheric $P(T)$ relation in hand, we calculate both $r_\phi$ and the mass coordinate $M$ of the photospheric layers, by integrating (\ref{eq:ktdrdM}) and (\ref{eq:ktdPdM}).

\subsection{Resulting stellar models}\label{sec:resultingStellarModels}

\begin{figure}
\centering
\subfloat[Throughout the star]{\label{fig:soundSpeed1}\includegraphics[width=0.46\textwidth]{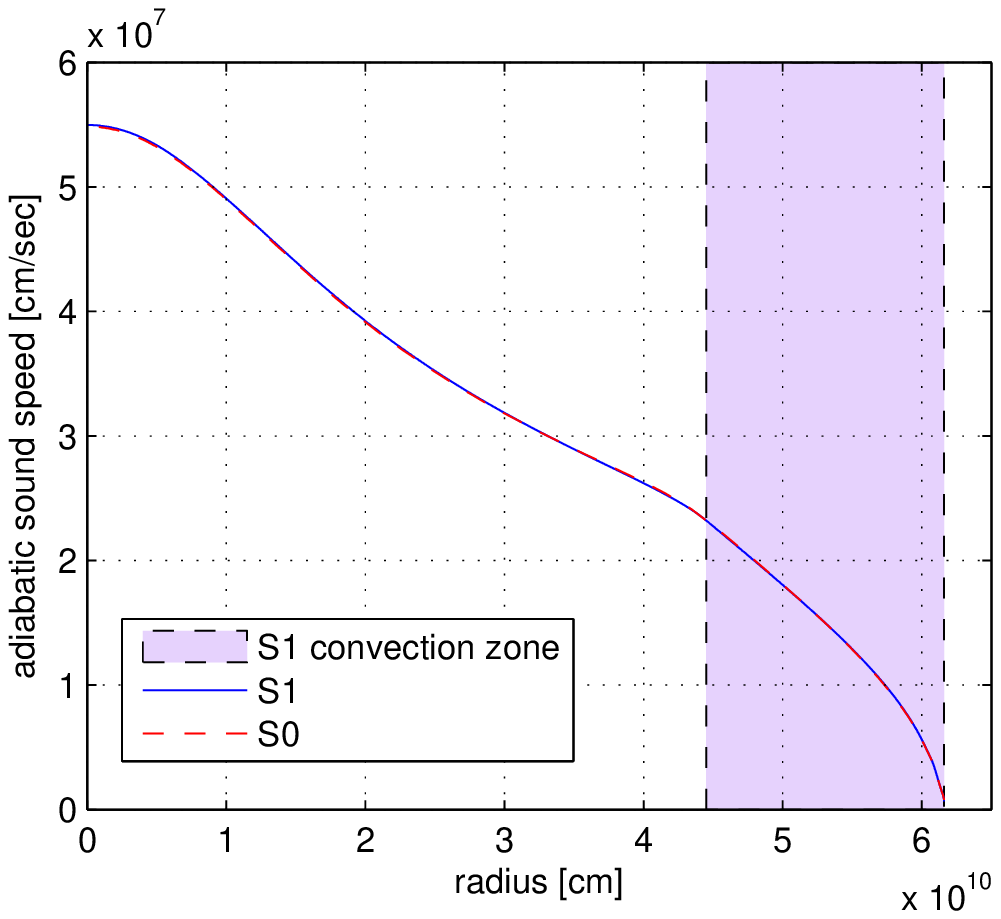}} \\
\subfloat[Near the surface]{\label{fig:soundSpeed2}\includegraphics[width=0.46\textwidth]{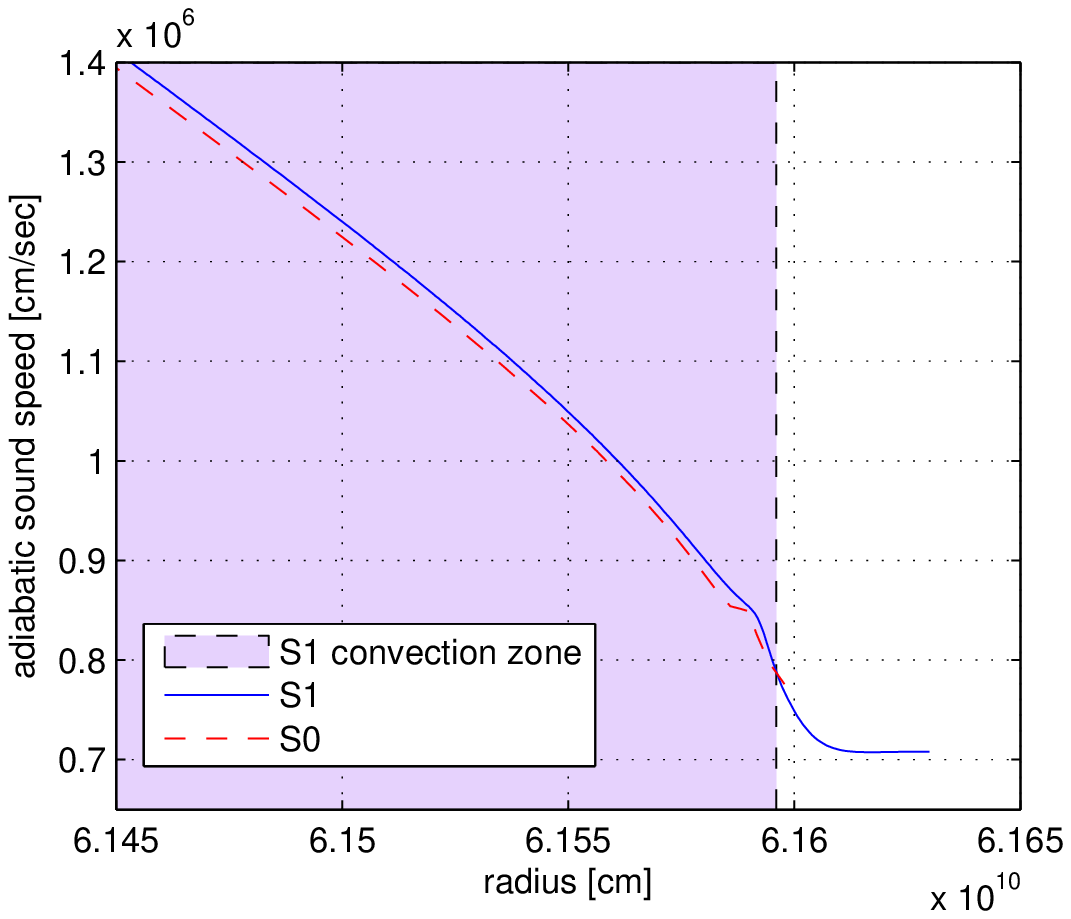}}
\caption{Adiabatic sound speed profiles of $S1$ and $S0$.}
\label{fig:soundSpeed}
\end{figure}

\begin{figure}
\includegraphics[width=0.46\textwidth]{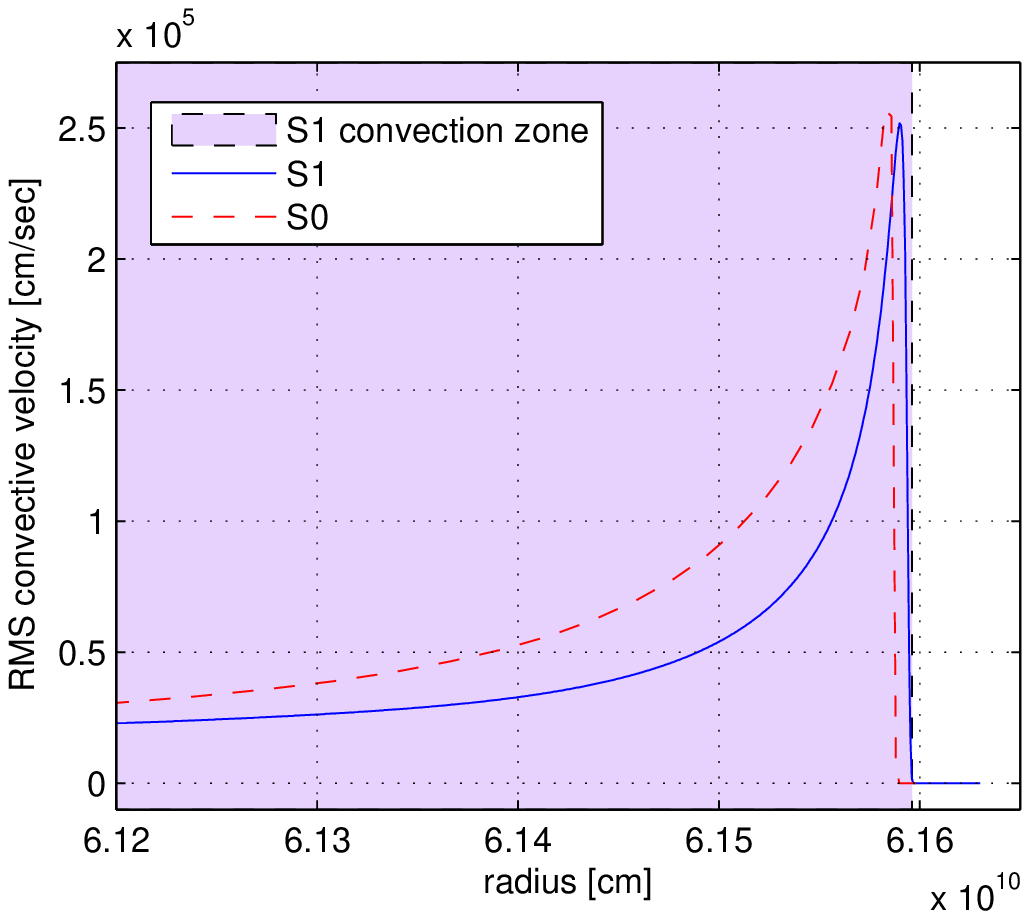}
\caption{RMS convective velocities near the upper boundary of the outer convection zone in $S1$ and $S0$.}
\label{fig:convection}
\end{figure}

Our zero-age models were constructed for the same composition ($X$ and $Z$), as was used by \citet{PolsTout1995} in computing a zero-age main-sequence. These values are slightly different from those presently found by standard solar models \citep[see e.g.][]{Turck2001}, and were mainly chosen for the purpose of concreteness. This choice also allowed us to compare the results of our stellar structure code with the $1 M_{\odot}$ model of \citet{PolsTout1995} (denoted $S0$), which is readily available online \citep[see][]{Townsend}. The value of the mixing length parameter $\alpha_{\textrm{m}} \equiv l_{\textrm{m}}/H_P$, was chosen so that the base-photospheric radius, $r_\mr{e}$, of our spherical model $S1$, matched the corresponding radius of $S0$. In Figs.\,\ref{fig:soundSpeed} and \ref{fig:convection}, we show the adiabatic sound speed $a \equiv \sqrt{\Gamma_1 P / \rho}$, and the convective velocities $\mean{v_\mr{con}^2}^{1/2}$ of models $S0$ and $S1$.

A set of 12 additional binary models were constructed for identical values of $M$,$X$,$Z$, and $\alpha_{\textrm{m}}$, at different values of the separation parameter $d$ (defined in \S\ref{sec:binaryPotential}). These values, and some of the resulting stellar parameters are summarized in tables \ref{tbl:modelConstants} and \ref{tbl:modelResultingParameters}.
\begin{table}
\begin{tabular}{c*{4}c}
$M_{\odot}\left[\mr{g}\right]$ & $M/M_{\odot}$ & $X$ & $Z$ & $\alpha_{\textrm{m}}$ \\
\hline
$1.9891\times10^{33}$ & $1$ & $0.70$ & $0.02$ & $2.219028(5)$
\end{tabular}
\caption{Values used in constructing models $S1$-$S13$}
\label{tbl:modelConstants}
\end{table}
\begin{table*}
\begin{tabular}{l||c*{5}c}
model & $\log P_\mr{c}\left[\mr{dyne/cm^2}\right]$ & $\log T_\mr{c}\left[\mr{K}\right]$ & $\log T_\mr{e}\left[\mr{K}\right]$ & $r_{\phi,\mr{e}}/R_{\odot}$ & $d/R_{\odot}$ & $r_{x,\mr{e}}/R_{\odot}$ \\
\hline
$S0$ & $17.15042$ & $7.12628$ & $3.74939$ & $0.88565$ & - & - \\
$S1$ & $17.15080$ & $7.12942$ & $3.75717$ & $0.88565$ & - & - \\
$S2$ & $17.15067$ & $7.12920$ & $3.75661$ & $0.88633$ & $2.5000$ & $0.8941$ \\
$S3$ & $17.15063$ & $7.12913$ & $3.75627$ & $0.88709$ & $2.2000$ & $0.8991$ \\
$S4$ & $17.15054$ & $7.12898$ & $3.75557$ & $0.88855$ & $1.9000$ & $0.9086$ \\
$S5$ & $17.15033$ & $7.12862$ & $3.75411$ & $0.89144$ & $1.6000$ & $0.9299$ \\
$S6$ & $17.15012$ & $7.12827$ & $3.75274$ & $0.89395$ & $1.4500$ & $0.9526$ \\
$S7$ & $17.14991$ & $7.12792$ & $3.75135$ & $0.89643$ & $1.3500$ & $0.9799$ \\
$S8$ & $17.14980$ & $7.12774$ & $3.75064$ & $0.89772$ & $1.3100$ & $0.9967$ \\
$S9$ & $17.14967$ & $7.12753$ & $3.74977$ & $0.89924$ & $1.2700$ & $1.0202$ \\
$S10$ & $17.14956$ & $7.12735$ & $3.74897$ & $0.90059$ & $1.2400$ & $1.0460$ \\
$S11$ & $17.14948$ & $7.12721$ & $3.74833$ & $0.90163$ & $1.2200$ & $1.0713$ \\
$S12$ & $17.14940$ & $7.12707$ & $3.74751$ & $0.90281$ & $1.2000$ & $1.1137$ \\
$S13$ & $17.14935$ & $7.12701$ & $3.74707$ & $0.90349$ & $1.1906$ & $1.1654$ \\
\end{tabular}
\caption{Resulting stellar model parameters.}
\label{tbl:modelResultingParameters}
\end{table*}
In table \ref{tbl:modelResultingParameters}, $r_{x,\mr{e}}$ is defined as the distance along the $x$ axis between the stellar centre and the base of the photosphere nearest $L_1$ (computed for the binary models). 

\begin{centering}
\begin{figure*}
\centering
\subfloat[$S9$]{\label{fig:soundSpeedXY1}\includegraphics[width=0.30\textwidth]{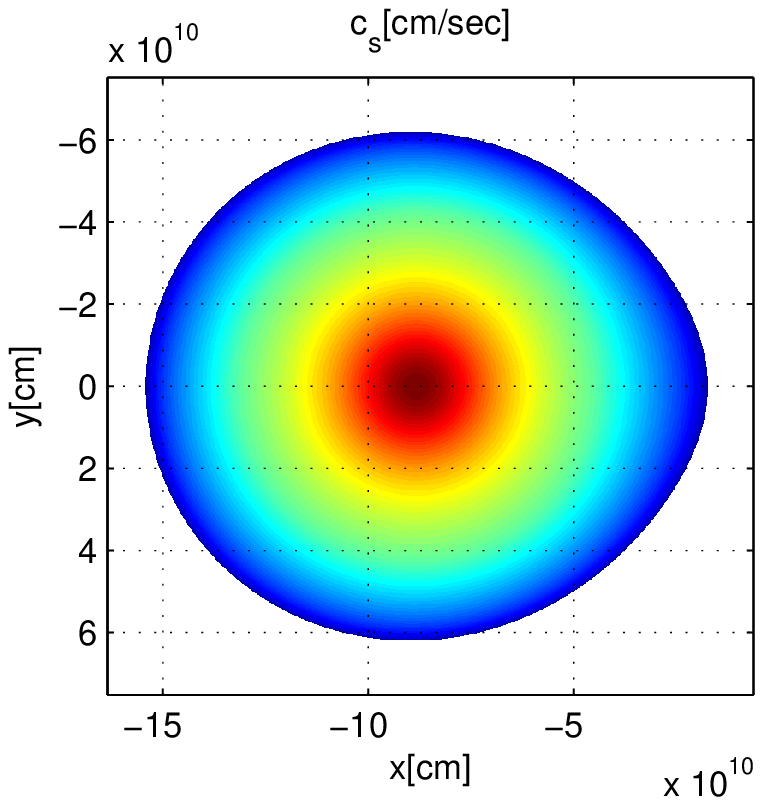}} 
\subfloat[$S12$]{\label{fig:soundSpeedXY2}\includegraphics[width=0.30\textwidth]{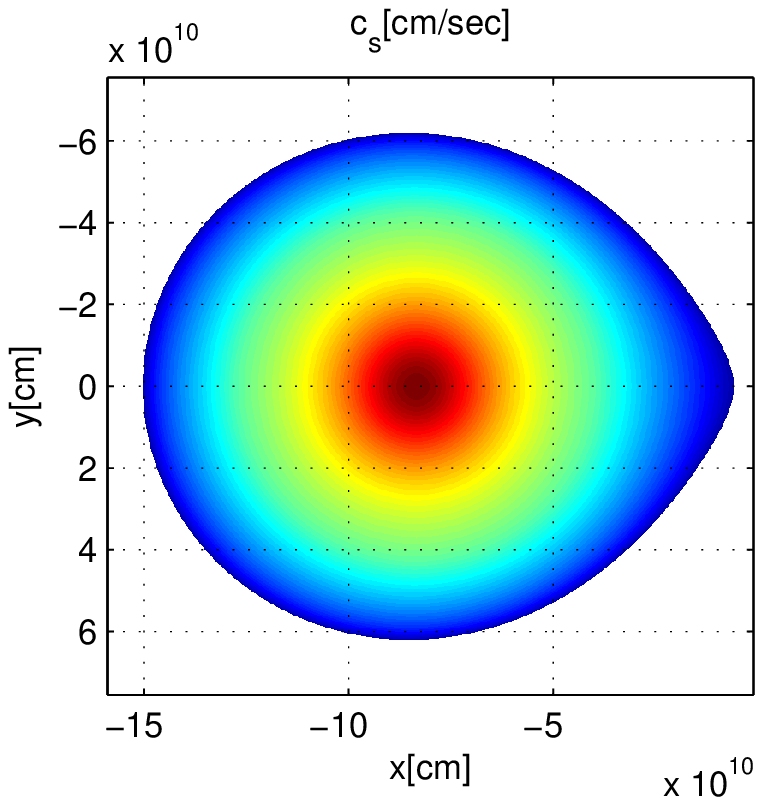}} 
\subfloat[$S13$]{\label{fig:soundSpeedXY3}\includegraphics[width=0.38\textwidth]{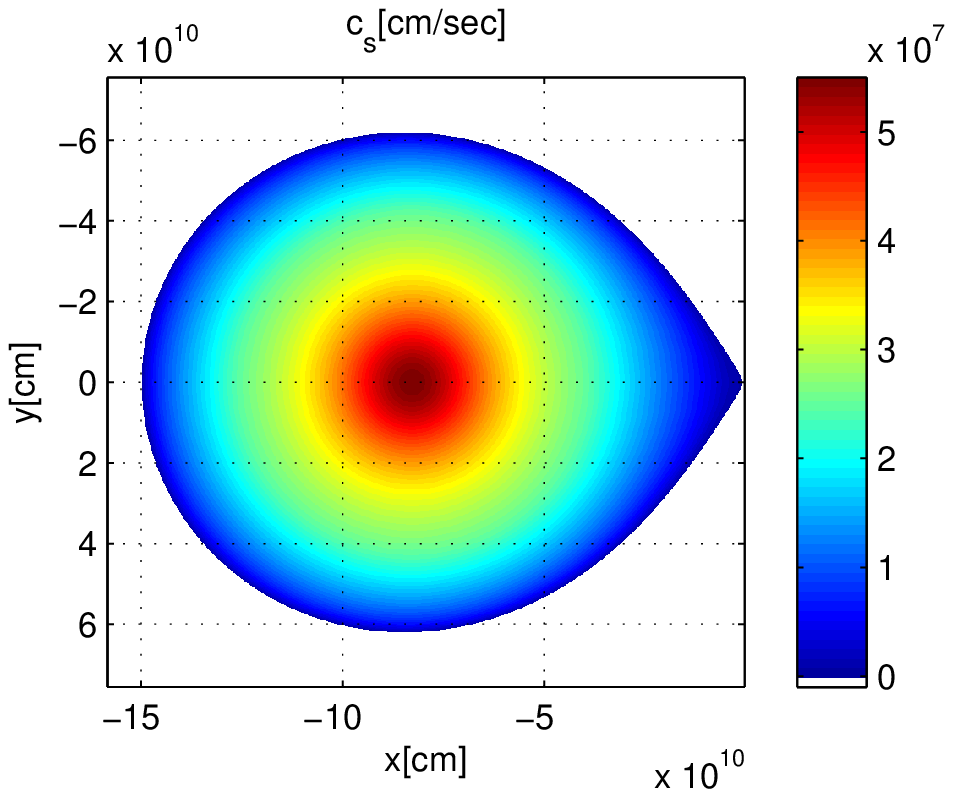}}
\caption{Adiabatic sound speed $a$ in three of the binary models in the $x$-$y$ plane.}
\label{fig:soundSpeedXY}
\end{figure*}
\end{centering}

\begin{figure*}
\centering
\subfloat[$S9$]{\label{fig:gravityXY1}\includegraphics[width=0.30\textwidth]{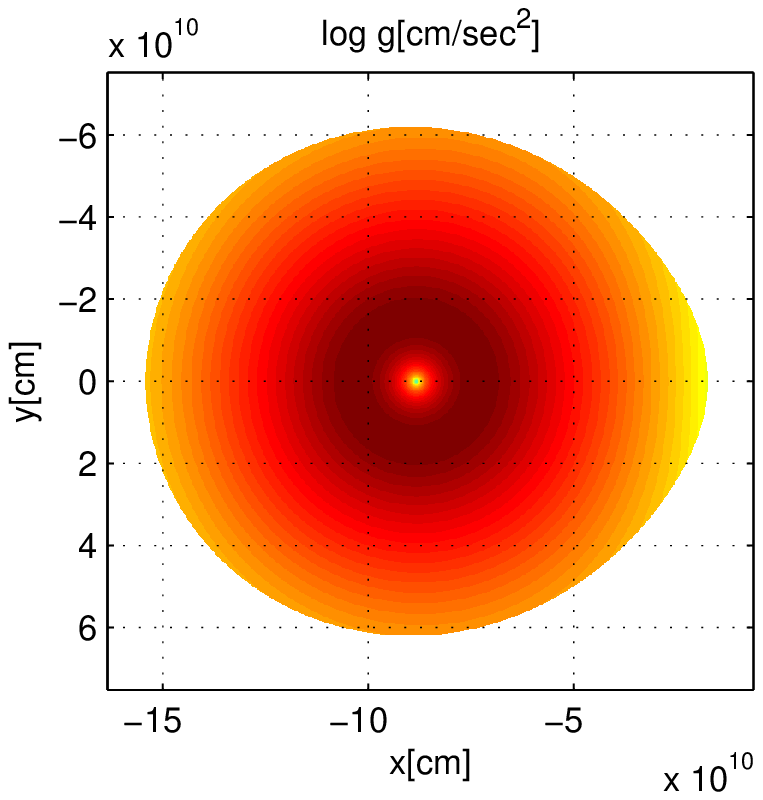}}
\subfloat[$S12$]{\label{fig:gravityXY2}\includegraphics[width=0.30\textwidth]{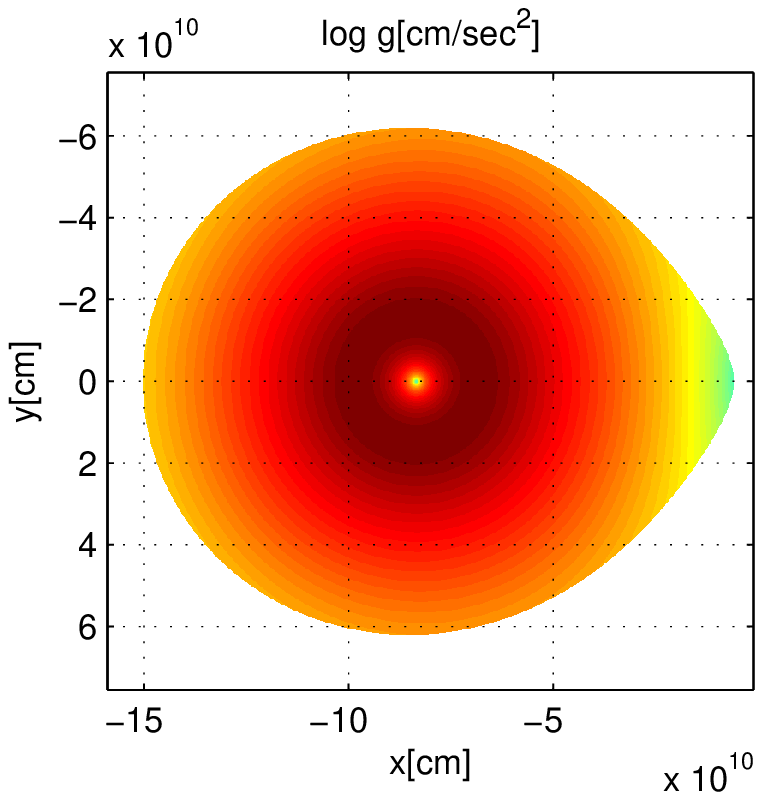}}
\subfloat[$S13$]{\label{fig:gravityXY3}\includegraphics[width=0.38\textwidth]{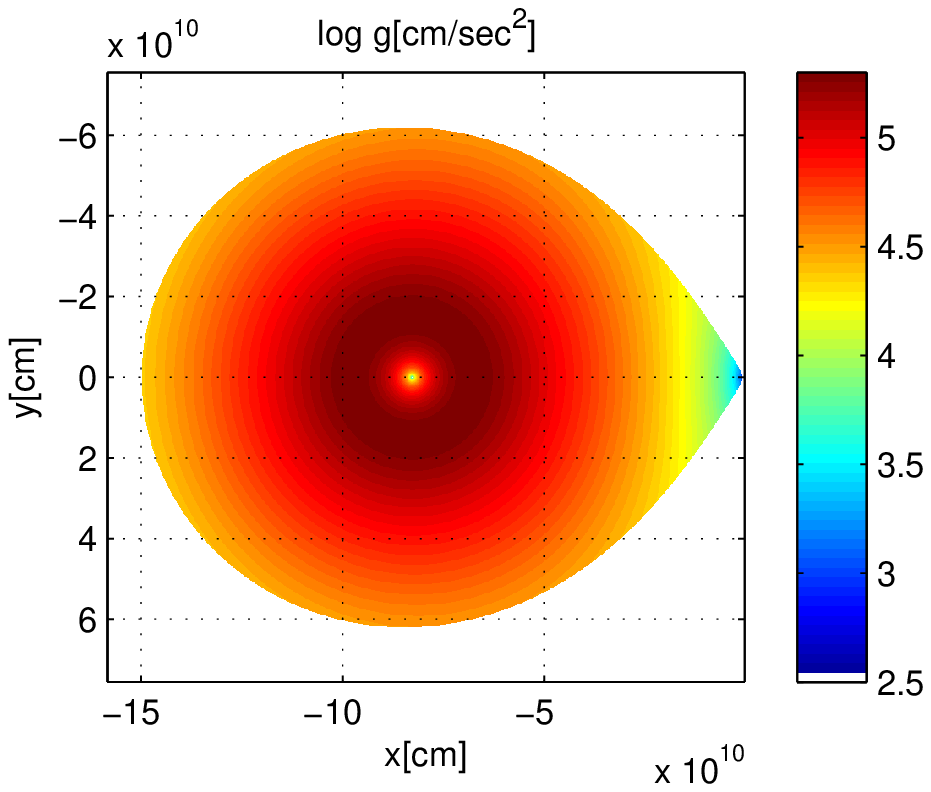}}
\caption{Effective gravity $g$ in three of the binary models in the $x$-$y$ plane.}
\label{fig:gravityXY}
\end{figure*}

In \S\ref{sec:propTrapping}, we have seen that the conditions at the surface, more specifically, the surface values of the effective gravity $g$, and the adiabatic sound speed $a$, determine the frequencies of the acoustic waves that may be trapped below the photosphere. The values of these quantities, in the $x$-$y$ plane, are shown in Figs.\,\ref{fig:soundSpeedXY} and \ref{fig:gravityXY}. Notice that $g$ varies on equipotentials, whereas $a$ does not. The increasing surface distortion of the binary models, for decreasing values of the separation parameter $d$, is also apparent in Figs.\,\ref{fig:soundSpeedXY}, \ref{fig:gravityXY}, and in the values of $r_{x,\mr{e}}(d)$ in table \ref{tbl:modelResultingParameters}.


\section{Online code}

Both the binary stellar structure code and the ray tracing code described in this work are readily available online at \url{http://www.phys.huji.ac.il/~springer/binseis}.


\footnotesize{ 
  \bibliographystyle{mn2e}
  \bibliography{binseis}
}


\label{lastpage}
\end{document}